\documentclass{SCIS2025}
\usepackage{cite}
\usepackage{amsmath,amssymb}
\usepackage{algorithm}
\usepackage{booktabs}
\usepackage{graphicx}
\usepackage{textcomp}
\usepackage{xcolor}
\usepackage{multicol}
\usepackage{multirow}
\usepackage{array}

\usepackage{diagbox}
\usepackage{slashbox}
\usepackage{color}
\usepackage{colortbl}
\usepackage{makecell}
\usepackage{latexsym}
\usepackage{caption}

\usepackage{caption}
\RequirePackage[pdfstartview=FitH,breaklinks,linkcolor=black,citecolor=black,filecolor=black,urlcolor=black,hyperindex,CJKbookmarks,
colorlinks=false,
pdfborder={0 0 1}
]{hyperref}

\hypersetup{
	linkbordercolor=red,
	citebordercolor=green,
	urlbordercolor=blue,
	filebordercolor=red
}

\begin{document}
\ArticleType{RESEARCH PAPER}
\Year{2025}
\Month{January}
\Vol{68}
\No{1}
\DOI{}
\ArtNo{}
\ReceiveDate{}
\ReviseDate{}
\AcceptDate{}
\OnlineDate{}
\AuthorMark{}
\AuthorCitation{}

\title{Does Movable Antenna Present A Dual-edged Nature? From the Perspective of Physical Layer Security: A Joint Design of Fixed-position Antenna and Movable Antenna
}{Yu K, Wang W X, Liu X W, et al. Does Movable Antenna Present A Dual-edged Nature? From the Perspective of Physical Layer Security: A Joint Design of Fixed-position Antenna and Movable Antenna}

\author[1,3]{Kan Yu}{}
\author[2]{Wenxu Wang}{}
\author[2]{Xiaowu Liu}{}
\author[1]{Yujia Zhao}{}
\author[1]{Qixun Zhang}{}
\author[1]{Zhiyong Feng}{}
\author[3]{Dong LI}{{dli@must.edu.mo}}


\address[1]{Key Laboratory of Universal Wireless Communications, Ministry of Education, \\
	 Beijing University of Posts and Telecommunications, Beijing 100876, P.R. China}
\address[2]{School of Computer Science, Qufu Normal University, Rizhao 276826, P.R. China}
\address[3]{School of Computer Science and Engineering, \\ Macau University of Science and Technology, Taipa, Macau 999078, P.R. China}
\abstract{
In conventional artificial noise (AN)-aided physical-layer security systems, fixed-position antenna (FPA) arrays exhibit inherent vulnerability to coverage gaps due to their static spatial configuration. Adversarial eavesdroppers can strategically exploit their mobility to infiltrate these spatial nulls of AN radiation patterns, thereby evading interference suppression and successfully intercepting the confidential communication. To overcome this limitation, in this paper, we investigate a hybrid antenna deployment framework integrating FPA arrays and movable antenna (MA) arrays (denoted by FMA co-design) to address the security performance in dynamic wireless environments, based on the fact that MA arrays enable channel reconfiguration through localized antenna repositioning, achieving more higher spatial degree of freedom (DoF). Enabled by FMA co-design framework, FPA arrays ensure baseline connectivity for legitimate links while MA arrays function as dynamic security enhancers, replacing conventional static AN generation. Furthermore, we formulate a non-convex optimization problem of the secrecy rate maximization through jointly optimizing MA positioning, FPA beamforming, and MA beamforming under practical constraints. the solution employs a dual-algorithm approach: Nesterov momentum-based projected gradient ascent (NMPGA) accelerates convergence in continuous position optimization, while alternating optimization (AO) handles coupled beamforming design. Experimental evaluations demonstrate that the proposed FMA co-design framework achieves significant secrecy performance gains over individual optimization benchmarks, yielding \textcolor{red}{42.34\% and 9.12\%} improvements in secrecy rate compared to isolated FPA for AN generation and MA for confidential information baselines, respectively. This work establishes a comprehensive solution for MA-FPA coexistence strategies in mission-critical mobile networks, balancing systems security, power efficiency, and mechanical practicality.

}


\maketitle

\section{Introduction}
Combining beamforming (BF) technology with fixed position antenna (FPA) arrays, where all antennas maintain a static spatial configuration, the artificial noise (AN) has emerged as a critical mechanism for enhancing wireless physical layer security. By concentrating interference signals specifically toward eavesdroppers (Eves) while preserving legitimate users' reception quality, AN significantly elevates Eves' decoding complexity and improves communication confidentiality. However, the intrinsic openness of wireless channels, coupled with the dynamic nature of mobile receivers and the stringent directional requirements of AN-BF, poses substantial challenges to secure information transmission. A critical vulnerability arises when directional AN-BF inadvertently targets legitimate mobile receivers while the BF of confidential signals aligns with Eves, which completely compromises security \cite{hong2020fixed}. Moreover, the fixed geometric constraints of FPA arrays impose inherent correlations/orthogonality among steering vectors across different angles, fundamentally limiting their capacities to achieve an adaptable interference nulling and a dynamic beam coverage \cite{tsai2014power}. These inherent limitations underscore the urgent need to address security vulnerabilities in conventional AN-BF implementations using FPA arrays.

To overcome these fundamental constraints of FPA arrays, the movable antenna (MA) technology has recently emerged as a paradigm-shifting solution. By enabling precise position and orientation adjustments flexibly through stepper motor control within a defined spatial region, MA systems provide an unprecedented spatial degree of freedom (DoF) and enhanced electromagnetic field manipulation capability \cite{tang2024secure,Zhu2024Modeling}. Unlike their FPA counterparts constrained by fixed steering vector relationships, MA arrays achieve a dynamic reconfigurability across one-dimensional (1D)-6D spatial dimensions (both position and orientation) \cite{Zhu2023Movable}. This adaptability allows a real-time optimization of steering vectors for multi-dimensional BF adjustments, effectively mitigating security risks associated with rigid AN-BF directionality in FPA systems. \emph{Importantly, FPA and MA technologies exhibit complementary advantages: FPAs ensure a stable coverage and low-latency operation through fixed position configurations, while MA provide enhanced spatial DoF through dynamic spatial adaptations, addressing coverage gaps.}  \textbf{This symbiotic relationship enables a coordinated security strategy: FPA arrays maintain a baseline connectivity while MA arrays actively disrupt eavesdropping channels through spatial signature randomization and adaptive interference targeting.}

The synergistic integration of FPA and MA technologies presents a transformative opportunity to enhance physical layer security through spatial resource orchestration. This integration raises fundamental research questions: \emph{\textbf{How can system designers jointly optimize the fixed-coverage advantages and directional BF capabilities of FPAs with the flexible DoF provided by MAs to achieve dual objectives:} 1) confidential signal enhancement for legitimate receivers, and 2) adaptive AN-BF interference against Eves?} More formally: \textbf{\emph{What mathematical framework enables the co-design of FPA/MA BF matrices and MA positioning configurations to maximize security performance?}} Current literature lacks comprehensive solutions to this critical problem, particularly regarding the spatial-temporal coordination between static and dynamic antenna systems in traditional physical layer security.

Building upon these foundational insights, we introduce a groundbreaking hybrid antenna architecture that orchestrates FPA and MA technologies through spatial-electromagnetic synergy for robust wireless security. The proposed architecture ensures a sustained confidential transmission via FPA arrays to the authorized receiver while dynamically deploying MA systems to generate spatially focused AN signals that actively disrupt multiple Eves. 
\emph{To our knowledge}, this work represents the first systematic investigation into the \textbf{FPA-MA co-design} for physical layer security enhancement, addressing critical gaps in existing wireless security paradigms. The principal innovations of this paper are summarized as follows.
\begin{itemize}
	\item Considering the high mobility scenario of the IoV, an innovative hybrid architecture is proposed that combines FPA and MA to enhance physical layer security. This architecture leverages the flexibility of MA to replace traditional AN techniques while maintaining basic communication services through FPA, effectively balancing system performance and security.
	\item A non-convex problem is formulated to maximize the confidential rate. The optimization problem is decomposed into three subproblems: MA position optimization, the BF of MA optimization, and the BF of FPA optimization, which are alternately optimized using the alternating optimization (AO) algorithm. To obtain a high-quality feasible solution, we propose an Nesterov momentum-based projected gradient ascent (NMPGA) algorithm to optimize the nonconvexity of the objective function.
	\item Through simulation experiments, the superiority of the proposed scheme is demonstrated in terms of convergence efficiency, secrecy rate performance and robustness. The results show that the proposed \textbf{FPA-MA co-design} framework achieves a better secrecy performance compared to both conventional AN-based and pure MA-based approaches, particularly in dynamic IoV scenarios with varying distances between legitimate users and eavesdroppers.
\end{itemize}

The remainder of the paper is organized as follows.  The related works are demonstrated in Section \ref{sec:related_works}. In Section \ref{sec:system_model}, we introduce the network model and perform a comprehensive problem analysis, along with presenting the necessary preliminaries. Section \ref{sec:optimization} is the detailed derivation process of the optimization problem and the details of the algorithm design. we describe the experiments in Section \ref{sec:numerical_results}. Finally, Section \ref{sec:conclusions} concludes the paper and our future works.
\begin{table*}[t]
	\small
	\caption{\small A brief summary of our proposed methods and current popular physical layer security methods}
	\centering
	\label{tab:methods}
	\begin{tabular}{|>{\centering\arraybackslash}m{1.5cm}|>{\centering\arraybackslash}m{4cm}|>{\centering\arraybackslash}m{0.6cm}|>{\centering\arraybackslash}m{0.6cm}|>{\centering\arraybackslash}m{1.0cm}|>{\centering\arraybackslash}m{5.8cm}|} \hline
		\textbf{Reference} & \textbf{Core ideas of physical layer security design} & \textbf{FPA} & \textbf{MA} & \textbf{perfect CSI} & \textbf{Summaries} \\ \hline
		\cite{Jiang2024Improving} & AN-based security through distributed jammers & $\checkmark$ & $\times$ & $\checkmark$ & Adaptive jamming and interference cancellation for eavesdropper channel degradation \\ \hline
		\cite{Zhu2016Improving} & SINR-based security with cooperative AN beamforming & $\checkmark$ & $\times$ & $\checkmark$ & SDR-based power minimization with SINR differentiation \\ \hline
		\cite{xu2021low}, \cite{li2019uav} & Mobile jamming with UAV-generated AN & $\checkmark$ & $\times$ & $\checkmark$ & UAV trajectory and power optimization for collaborative jamming \\ \hline
		\cite{Hu2024Secure}, \cite{Xiong2025Analog}  & Secrecy rate maximization via MA position flexibility & $\times$ & $\checkmark$ & $\checkmark$ & Joint antenna position and beamforming optimization for secrecy rate \\ \hline
		\cite{wu2023Discrete} & Discrete MA position selection for secured QoS & $\times$ & $\checkmark$ & $\checkmark$ & Power minimization with QoS guarantees via discrete MA positioning \\ \hline
		\cite{Ding2025FD} & Full-duplex security with 2D MA spatial diversity & $\times$ & $\checkmark$ & $\checkmark$ & Joint multi-parameter optimization against collaborative eavesdropping \\ \hline
		\cite{yan2025movableantennaaidedmultiuser} & Dual time-scale security optimization for dynamic channels & $\times$ & $\checkmark$ & $\times$ & Dual time-scale optimization balancing security and energy efficiency \\ \hline
		\cite{feng2024movable} & Security under imperfect eavesdropper CSI with virtual MA modeling & $\times$ & $\checkmark$ & $\times$ & Virtual MA modeling for security under imperfect eavesdropper CSI \\ \hline
		\cite{ma2024movableantennaaidedsecuretransmission} & Multi-technology security fusion with MA-RIS-ISAC & $\times$ & $\checkmark$ & $\checkmark$ & MA-RIS-ISAC integration for enhanced secrecy performance \\ \hline
		\textbf{methods in this paper} & \textbf{Hybrid FPA-MA co-design deployment} & $\checkmark$ & $\checkmark$ & $\checkmark$ & NMPGA and AO-based hybrid FPA-MA optimization for secure IoV communications \\ \hline
	\end{tabular}
	\\
	\vspace{0.5cm} 
	\raggedright 
	\textbf{Note:} CSI: channel state information; ISAC: integrated sensing and communications.
\end{table*}

\section{Related Works} \label{sec:related_works}
The physical layer security methods are widely studied in IoV networks due to significance of driving safety. In IoV communication scenarios, conventional physical layer security approaches, such as AN \cite{wang2020physical}, reconfigurable intelligent surface (RIS)\cite{Deng2024Reconfigurable}, and unmanned aerial vehicle (UAV)\cite{Yu2024SuRLLC}, have been extensively studied and proven effective in enhancing communication security. In this section, we provide a comprehensive review of the current popular physical layer security methods, and summarize them in Table \ref{tab:methods} 
from the perspective of design ideas, advantages and limitations. 

\subsection{FPA empowered physical layer security Methods}
AN technology, as one of the quintessential approaches in physical layer security schemes, has emerged as a highly effective mechanism for degrading eavesdropping channels.
By generating AN to obfuscate confidential signals overheard by Eves, the discrepancy between the channels of legitimate users and Eves is effectively widened.
In \cite{Jiang2024Improving}, Jiang \textit{et al.}  investigated a distributed antenna system (DAS) where a base station (BS) equipped with multiple distributed antennas communicates with a legitimate user while multiple non-colluding Eves attempt to intercept the transmission. They employed a friendly jammer to emit AN signal, creating intentional interference to confuse and degrade the eavesdroppers' channel quality. The adaptive jamming (AJ) scheme and the interference cancellation (IC) scheme were designed to enhance interference management. 

Considering AN-BF design and power allocation between legitimate signals and AN signals, Hu \textit{et al.} emphasized the importance of attracting AN to physical layer security \cite{hu2019minimization}. 
Zhu \textit{et al.} \cite{Zhu2016Improving} designed a SINR-based cooperative beamforming strategy where AN with BF is employed to enhance the physical layer security of both primary users (PUs) and secondary users (SUs). The authors converted the initially non-convex problem into a convex semi-definite programming (SDP) problem using semi-definite relaxation (SDR) techniques.
Furthermore, considering the role of AN-BF in characterizing the trade-off between reliability and security, in \cite{wang2020physical}, Wang \textit{et al.} introduced the concept of effective secrecy throughput to quantify the average data rate of secure transmission of confidential information.  Similar works have also been done in \cite{Jiang2021Secrecy}, \cite{Liu2022Throughput}, and \cite{Li2023Multi}. 
In addition, AN can be combined with the UAV as a jammer to achieve collaborative safety by jointly optimizing the UAV trajectory and transmission power\cite{xu2021low}, \cite{li2019uav}.

Despite the progress of FPA-based AN solutions, these methods cannot be directly extended to IoV networks due to the constraints of the fixed geometry of FPAs.
1) The static nature of FPAs limits their ability to adapt to rapidly changing channel conditions caused by high-speed vehicle movement, leading to potential beam misalignment and reduced interference efficiency. 2) Predictable antenna locations make the system vulnerable to sophisticated eavesdropping strategies, especially in scenarios where attackers can exploit known antenna positions and fixed AN patterns. 3) Maintaining effective AN coverage while optimizing power allocation becomes challenging under the strict latency requirements of vehicular networks. These mean that it cannot accurately and timely adjust the channel to adapt to the latest communication scenarios, and spatial flexibility is limited \cite{zeng2024multi}.

\subsection{Physical Layer Security based on Joint Design of Antenna Positions and BF Matrix}
\subsubsection{Physical layer security methods under the assumption of perfect CSI}
In terms of MA, it offers a higher spatial DoF and flexible BF design, which has been regarded as a technology with great potential for wireless physical layer security. In \cite{Xiong2025Analog} and \cite{Hu2024Secure}, the authors considered scenarios involving a single Eve and multiple colluding eavesdroppers, respectively, deploying one-dimensional region-constrained MAs at the transmitter side. These works maximized secrecy rate through joint optimization of antenna positions and beamforming matrices. Meanwhile, \cite{wang2024movableantennaarrayaided} and \cite{Liu2025Covert} extended these concepts to covert communication scenarios, addressing communication covertness maximization problems under single and multiple Willies scenarios, respectively. As scenario complexity increases, MA dimensionality has progressively expanded to 2D/3D implementations. Cheng \textit{et al.} jointly optimized the 2D MA position and beamforming matrix to minimize power consumption and maximize confidentiality rate in eavesdropping scenarios, fully utilizing the DoF of MA two-dimensional space and improving the security performance of the system. Wu \textit{et al.} \cite{wu2023Discrete} modeled MA movement as a discrete optimization problem, minimizing total transmission power while guaranteeing users' quality of service (QoS). Ding \textit{et al.} \cite{Ding2025FD} proposed a full-duplex multi-user communication architecture that effectively counters collaborative interception by multiple Eves through joint optimization of 2D MA positions, base station's transmit and receive beamformers, artificial noise beamformers, and uplink power, resulting in significant improvements in sum secrecy rate (SSR). Notably, most of these studies assume perfect CSI, with implementation dependent on efficient channel estimation methods. Zhu \textit{et al.} \cite{Ma2023Compressed} proposed a continuous transmitter-receiver compressed sensing (STRCS) channel estimation framework that significantly reduces pilot overhead. Xiao \textit{et al.} \cite{Xiao2024Channel} further leveraged multipath sparsity to construct statistical channel models, providing theoretical support for high-dimensional MA system deployment.

\subsubsection{Physical layer security methods under the assumption of imperfect CSI}
To address the overhead associated with real-time CSI acquisition, research focus has shifted toward statistical and robust optimization strategies. Yan \textit{et al.} \cite{yan2025movableantennaaidedmultiuser} proposed a dual time-scale optimization framework that combines long-term MA position planning based on statistical CSI with dynamic precoding matrix adjustments based on instantaneous CSI, balancing security and energy efficiency in mobile user scenarios. In our previous work \cite{feng2024movable}, we introduced the concept of virtual MAs to model imperfect CSI for Eves, ensuring system confidentiality through joint beamforming and MA position optimization when eavesdropper CSI is unknown or partially known. Hu \textit{et al.} \cite{Hu2024TMC} derived closed-form expressions for secrecy outage probability under Rician fading when only the statistical line-of-sight (LoS) component of the eavesdropping channel is available, and proposed an alternating projected gradient ascent (APGA) algorithm to optimize MA positions, addressing the problem of minimizing secrecy outage probability. Furthermore, the integration of MAs with emerging technologies further enhances physical layer security performance. Ma \textit{et al.} \cite{ma2024movableantennaaidedsecuretransmission} achieved significant secrecy rate improvements in RIS-assisted ISAC systems through joint optimization of MA positions, beamforming, and intelligent surface phases. Xie \textit{et al.} \cite{Xie2024CovertPlusRIS} combined deep reinforcement learning (DRL) for MA trajectory optimization, achieving higher covertness rates in multi-Willies monitoring scenarios.

While existing studies have demonstrated the significant potential of MA in physical layer security communications, most of these investigations assume that the latency of antenna movement is negligible or tolerable. These pioneering works have established solid theoretical foundations and optimization frameworks for MA-based secure communications, particularly excelling in scenarios with static or low-mobility users where the antenna movement delay has minimal impact on system performance. However, the convergence speed of existing optimization algorithms cannot meet the ultra-low latency requirements in vehicular networks. On one hand, the relative movement between MA and high-speed vehicles compounds the complexity of channel variations, further increasing the probability of beam misalignment errors. On the other hand, the mechanical limitations of MA mobility and the frequent channel estimation impact delay and energy consumption.

\subsection{The difference between our work and previous studies}
Based on the existing research proposals, we combined the advantages of FPA and MA to investigate the FPA-MA system. The difference between our work and previous studies is described as follows:
\begin{itemize}
	\item Differing from traditional physical layer security schemes that rely on FPA arrays for AN generation \cite{wang2020physical,hu2019minimization,Jiang2021Secrecy,Liu2022Throughput, Li2023Multi},our proposed FPA-MA co-design framework adds more additional spatial DoF provided by the MA, which can generate more effective interference to the Eve and adapt more effectively to rapid channel variations, thereby improving secrecy performance under dynamic communication conditions.
	\item In contrast to existing MA-assisted security methods that utilize only movable antennas \cite{Cheng2024Enabling, Hu2024Secure, feng2024movable, mei2024movable, tang2024secure}, our hybrid FPA-MA architecture explicitly addresses the latency and coverage challenges in high-mobility IoV scenarios. By leveraging the FPA component for stable, low-latency communication coverage, the system maintains reliable baseline service, while simultaneously exploiting the MA's mobility to introduce adaptive interference for security enhancement. This dual-antenna design enables improved physical-layer security without sacrificing communication reliability, even as vehicles move at high speeds and network conditions change rapidly.
\end{itemize}

\emph{Notations:} In this article, $\textbf{x}^T$ and $\textbf{x}^H$ represent the transpose and conjugate transpose of the matrix or vector $\textbf{x}$, respectively. $ \mathbb{C}^{a \times b}$ and $ \mathbb{R}^{a \times b}$ respectively denote \(a \times b\) dimensional complex matrices and \(a \times b\) dimensional real matrices. \(||\mathbf{x}||_{2}\) represents the 2-norm of the vector \(\mathbf{x}\). \(\text{tr}(\mathbf{x})\) and \(\rm{diag}(\mathbf{x})\) respectively represent the trace and the diagonal matrix with diagonal elements \(\mathbf{x}\). \(\left| \mathbf{x} \right|\) represents taking the modulus of vector \(\mathbf{x}\). \(\nabla_{x}\) and \(\frac\partial{\partial x}\) respectively denote the gradient operator and the partial derivative operator. \([x]^{+}\) represents \(\max\{x, 0\}\).

\section{System Model And Problem Formulation} \label{sec:system_model}
In this section, we present the system model for secure MA-assisted wireless communication in the context of IoV. The channel models of FPA-Bob/Eves and MA-Bob/Eves are then analyzed, respectively. Finally, we formulate an optimization problem
to maximize the secrecy rate.

\subsection{System Model}
\begin{figure}[htbp]
	\centerline{\includegraphics[width=0.7\linewidth]{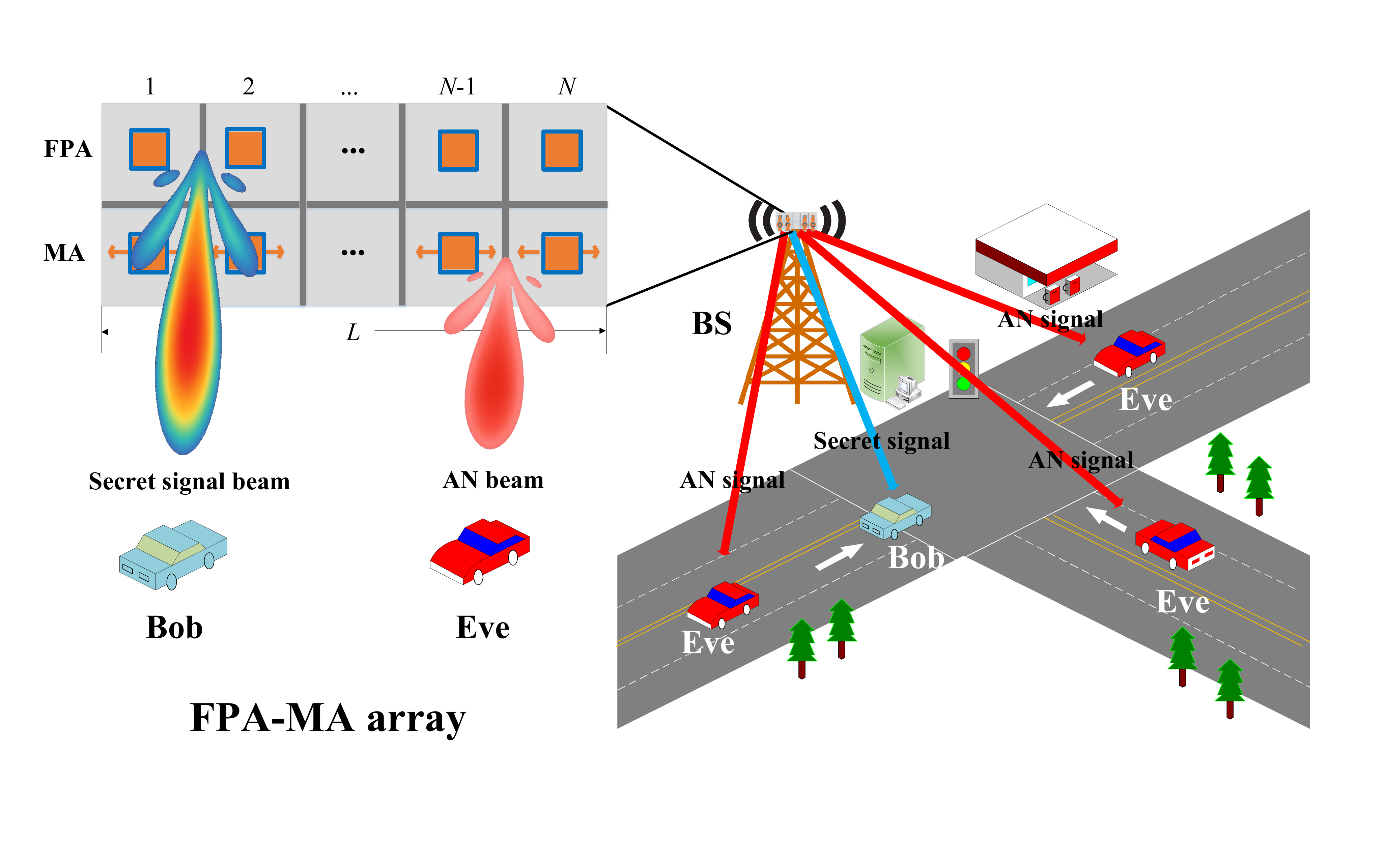}}
	\caption{\small Cooperative FPA-MA enabled communication model}
	\label{fig:Network Model}
\end{figure}

As illustrated in Fig. \ref{fig:Network Model}, the BS transmits the confidential information to the Bob, in the presence of $M$ Eves. To ensure the communication reliability and security simultaneously, the BS is equipped with two kinds of antennas with $N$, respectively, namely the FPA and the MA. In detail, the FPA is utilized to design the BF matrix of the confidential information, while the MA is used to generate AN signals with the BF design for pointing to the Eves, since the positions of the MA can be more flexibly adjusted and have a higher level of spatial DoF. These two types of antennas are deployed in a linear array, as shown in Fig. \ref{fig:Channel Model}.
For the FPA, the position vector is $\mathbf{x}_{\text{FPA}} = \left[ 0, d_{\min}, \cdots, (N-1)d_{\min}\right]^{T}$, where $d_{\min}$ is the inter-element spacing (typically equals to a half wavelength).

To establish a tractable framework for analyzing the physical layer security supported by the MA systems, we make a simplified approximation assumption to model the positions of the MA in different time slots as follows: 
given a continuous moving duration $T_{\rm total}$ of the MA, it can be divided into $Q$ equal time slots with the length of $T_{\rm total}/Q$, and the duration of each time slot is sufficiently small. Consequently, within the duration of $T_{\rm total}/Q$, each of $N$ antennas can be approximately stationary, and corresponding CSI is constant. As a result, in the $t$-th time slot, the position of the $n$-th antenna is represented by $x_{n}[t]$ ($1 \leq n \leq N$, $1 \leq t \leq Q$), and the positions of all $N$ antennas can be expressed as $\mathbf{x}_\text{MA}[t] = [x_{1}[t], x_{2}[t], \cdots, x_{N}[t]]^{T} \in \mathbb{R}^{N \times 1}$. All antenna elements are constrained to move within the range of $[0,~L]$, where $L$ denotes the maximum length of the FPA-MA array. 

\subsection{Channel Model}
\begin{figure}[htbp]
	\centering
	\includegraphics[width=0.65\linewidth]{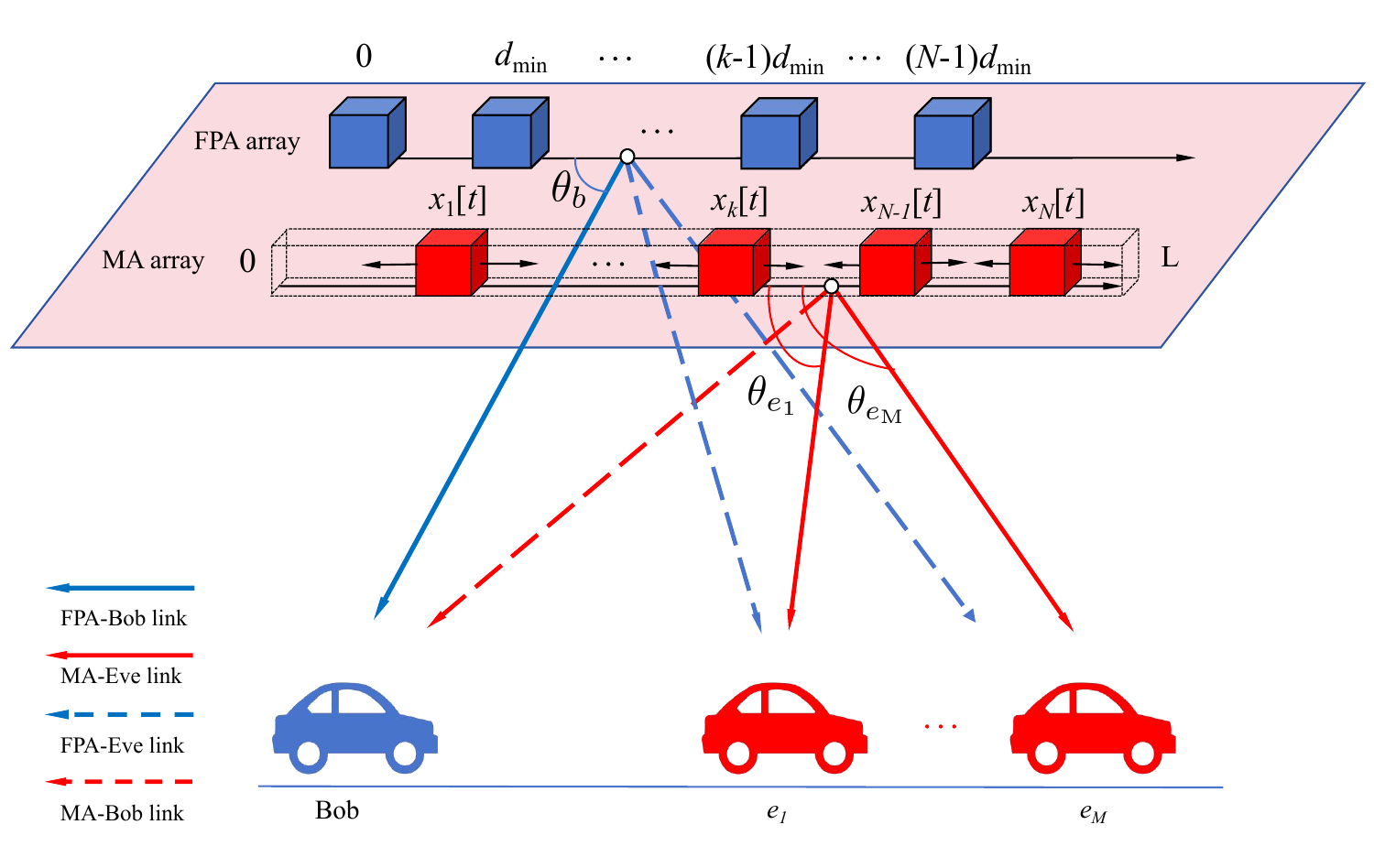}
	\caption{\small Steering angles for Bob and $i$-th Eve in FPA-MA array}
	\label{fig:Channel Model}
\end{figure}

To accurately characterize the wireless propagation characteristics between the BS and the receivers (Bob and Eves), the channel model can be established as follows. Let $\theta_{b}$ and $\theta_{e_i}$ represent the steering angle for the Bob and the $i$-th Eve, as shown in Fig. \ref{fig:Channel Model}. Similarly, let $d_{\rm BS, bob}$ and $d_{\rm BS, \it e_i}$ denote the distance between the BS and the Bob or the $i$-th Eve. Given the fixed positions $\mathbf{x}_{\rm FPA}$ of the FPA array and the steering angle $\theta$, the corresponding steering vector of the FPA array between the BS and the Bob as well as the $i$-th Eve can be represented as 
\begin{equation}
	\mathbf{a}(\mathbf{x}_{\text{FPA}}, \theta)=\sqrt{\frac{\kappa_{0}}{d^{\alpha}}}\left[e^{0}, e^{j \frac{2 \pi}{\lambda} d_{\min} \cos \theta}, \cdots, e^{j \frac{2 \pi}{\lambda} (N-1)d_{\min} \cos \theta}\right]^T,
\end{equation}
where $\theta = \{ \theta_{b}, \theta_{e_i} \}$, $d=\{d_b, d_{e_i}\}$, $i \in [1,2, \cdots, M]$, $\alpha$ is the path loss exponent, $\kappa_{0}$ signifies the reference path loss at $1$(m)\footnote{The reference path loss (W) is computed using the Free-Space Path Loss (FSPL) model, which posits ideal signal propagation in an unobstructed environment. The channel power at the reference distance $d_0$ is expressed as: $\kappa_{0}(d_0)=(\frac{\lambda}{4\pi d_0})^2$ \cite{Erceg1999pathlossmodel}.}, $\lambda$ represents the wavelength, and $[\cdot]^T$ denotes the transpose operation.
Let $$\mathbf{w}_\text{FPA}[t]=\{w_{\text{\rm FPA}}^1[t],w_{\text{\rm FPA}}^2[t],...,\\w_{\text{\rm FPA}}^N[t]\} \in \mathbb{C}^{N \times 1}$$ be the digital transmitting BF vector of the confidential information at the BS supported by the FPA.
The beam gain for the FPA array is then represented as: 
\begin{equation} \label{eq:4}
	G_\text{FPA}(\theta)= \left| \mathbf{a}^H(\mathbf{x}_\text{FPA},\theta) \mathbf{w}_\text{FPA}[t] \right|^2, \theta \in [0, \pi),
\end{equation}
where $H$ represents the conjugate transpose operation.

Different from conventional FPA arrays that employ static antenna spatial configurations of, MA arrays feature dynamic reconfigurability through position-adaptive movements, necessitating a fundamentally different approach to steering vector characterization. Building upon the theoretical framework established in \cite{Ma2024Multi}, the steering vector formulation for the MA array must incorporate both the time-varying antenna positions $\mathbf{x}_{\rm MA}[t]$ and the steering angle $\theta$. The complete mathematical representation can be expressed as
\begin{equation}\label{eq:steering vector MA}
	\mathbf{a}(\mathbf{x}_\text{MA}[t], \theta)= 
	\sqrt{\frac{\kappa_{0}}{ d^{\alpha}}} \left[e^{j \frac{2 \pi}{\lambda} x_1[t] \cos \theta}, e^{j \frac{2 \pi}{\lambda} x_2[t] \cos \theta}, \cdots, e^{j \frac{2 \pi}{\lambda} x_N[t] \cos \theta}\right]^T.
\end{equation}
As a result, the beam gain for  given MA array can be expressed as 
\begin{equation} \label{eq:2}
	G_{\text{MA}}(\theta)= \left| \mathbf{a}^H(\mathbf{x}_\text{MA}[t],\theta) \mathbf{w}_\text{MA}[t] \right|^2, \theta \in [0, \pi),
\end{equation}
where $\mathbf{w}_{\text{MA}}[t]=\{w_{\text{\rm MA}}^1[t], w_{\text{\rm MA}}^2[t],...,w_{\text{\rm MA}}^N[t]\}$ represents the BF vector for the MA array.

To sum up, when the BS transmits the confidential information with the FPA array and the AN signals with the MA array, respectively, the signal received at the Bob from the BS can be represented as
\begin{equation}\label{eq:receive_signal_bob}
	y_b = \mathbf{a}^H(\mathbf{x}_\text{FPA},\theta_b) \mathbf{w}_\text{FPA}[t]s_{\rm FPA} + \mathbf{a}^H(\mathbf{x}_\text{MA}[t],\theta_b) \mathbf{w}_\text{MA}[t]s_{\rm MA} + n_b,
\end{equation}
where $s_{\rm FPA}$ and $s_{\rm MA}$ represent signals send by FPA array and MA array, $n_b \sim \mathcal{CN}(0, \sigma^2)$ represents the Additive White Gaussian Noise (AWGN) with zero mean and average power $\sigma^2$ at the Bob.

Similarly, for the Eves, we consider the worst case that all Eves collaborate to intercept the confidential information. For this case, the aggregate signals received at the colluding Eve, denoted by $e^*$, can be mathematically represented as
\begin{align}\label{eq:receive_signal_eve}
	y_{e^*} =\underbrace{\sum_{i=1}^M \mathbf{a}^H(\mathbf{x}_\text{FPA},\theta_{e_i}) \mathbf{w}_\text{FPA}[t]s_{\rm FPA}}_{\text{secret signal send by FPA}} + \underbrace{\sum_{i=1}^M \mathbf{a}^H(\mathbf{x}_\text{MA}[t],\theta_{e_i}) \mathbf{w}_\text{MA}[t]s_{\rm MA}}_{\text{AN signal generated by MA}} + n_e,
\end{align}
where $n_e$ denotes the AWGN at the Eves and $n_e \sim \mathcal{CN}(0, \sigma^2)$.

Based on expressions of Eq. \eqref{eq:4}, Eq. \eqref{eq:2}, and Eq. \eqref{eq:receive_signal_bob}, the received signal-plus-noise ratio (SINR) of the confidential information at the Bob and the collusive Eve can be expressed as
\begin{equation}\label{eq:sinr bob}
	\text{SINR}_{b}[t] =\frac{\left|\mathbf{a}^H(\mathbf{x}_\text{FPA},\theta_b) \mathbf{w}_\text{FPA}[t] \right|^2}{\left|\mathbf{a}^H(\mathbf{x}_\text{MA}[t],\theta_b) \mathbf{w}_\text{MA}[t]\right|^2 + \sigma^2},
\end{equation}
and
\begin{equation}\label{eq:sinr eve}
	\text{SINR}_{e^*}[t] = \frac{\sum_{i=1}^M\left| \mathbf{a}^H(\mathbf{x}_\text{FPA},\theta_{e_i}) \mathbf{w}_\text{FPA}[t]\right|^2}{\sum_{i=1}^M \left| \mathbf{a}^H(\mathbf{x}_\text{MA}[t],\theta_{e_i}) \mathbf{w}_\text{MA}[t] \right|^2 + \sigma^2}.
\end{equation}
\subsection{Performance Metric: Average Achievable Secrecy Rate}
Based on the conclusions of \cite{Jiancheng2023Fundamental}, the achievable rate (bps/Hz) of the confidential information between the Alice and the Bob in the $t$-th time slot is $R_{\rm bob}[t] =\log_{2}(1+\text{SINR}_{b}[t])$. Similarly, the achievable rate (bps/Hz) between the Alice and the collusive Eve for decoding the confidential information in the $t$-th time slot is given by $R_{\rm eve}[t]=\log_{2}(1+\text{SINR}_{e^*}[t])$.
Hence, the achievable secrecy rate between the Alice and the collusive Eve in the $t$-th time slot can be represented as
\begin{displaymath}
	R_{\rm sec}[t]=\left[R_{\rm bob}[t]-R_{\rm eve}[t]\right]^+,
\end{displaymath}
where $[x]^+\triangleq \max(x,0)$, and the average secrecy rate of the system over all time slots is given by 
\begin{displaymath}
	\bar{R}_{\rm sec}=1/Q\cdot \sum_{t=1}^QR_{\rm sec}[t],
\end{displaymath}
which represents the long-term secrecy performance of the communication system across all time slots and provides a more comprehensive metric for evaluating the overall secrecy performance rather than focusing on instantaneous state.

\subsection {Optimization Problem Formulation}
Different from the conventional physical layer security method design, this paper aims to develop a novel physical layer security method by utilizing the strong directionality BF of the FPA array and flexible spatial freedom of the MA array, while satisfying the constraints on the maximum transmission power allocation of the confidential information and AN signals. That is, the antenna positions $\mathbf{x_\text{\rm MA}}[t]$ of the MA array, BF matrix $\mathbf{w}_\text{\rm FPA}[t]$ of the FPA array for transmitting the confidential information, and BF matrix $\mathbf{w}_\text{\rm MA}[t]$ of the MA array for transmitting AN signals are jointly optimized to maximize the average secrecy rate over all time slots. The optimization problem can be formulated as 
\begin{subequations}\label{eq:P1}
	\begin{align}
		\text{(P1):}& \max_{\mathbf{x}_\text{MA}[t], \mathbf{w}_{\text{MA}}[t], \mathbf{w}_{\text{FPA}}[t]} 1/Q\cdot\sum_{t=1}^Q \left[R_{\rm bob}[t]-R_{\rm eve}[t]\right]^+ \label{eq:6a}\\
		\text{s.t.} ~~ 
		& \left| x_i[t] - x_j[t] \right| \geq d_{\textrm{min}},1 \leq i < j \leq N , \label{eq:st1}\\
		& \left\{ x_s[t] \right\}_{s=1}^N \in [0,L] ,\label{eq:st2}\\
		& ||\mathbf{w}_{\text{MA}}[t]||_2^2\leq P_{\text{MA}} ,\label{eq:st3}\\
		& ||\mathbf{w}_{\text{FPA}}[t]||_2^2\leq P_{\text{FPA}},\label{eq:st4}
	\end{align} 
\end{subequations}
where constraint \eqref{eq:st1} ensures that at any time $t$, the minimum distance between any two antenna elements $x_i[t]$ and $x_j[t]$ of MA ($1 \leq i < j \leq N $) is not less than $d_{min}$, thereby preventing unwanted coupling effects between elements. The range of $\left\{x_s[t] \right\}_{s=1}^N$ in Eq. \eqref{eq:st2} represents the maximum distance, denoted by $L$, that each antenna can move potentially. $P_{\rm MA}$ and $P_{\rm FPA}$ in Eq. \eqref{eq:st3} and Eq. \eqref{eq:st4} are the maximum transmission power of the MA and the FPA, respectively \cite{Jia2024Physical}. 

However, Problem \eqref{eq:P1} is non-convex optimization problem, which is difficult to be solved directly, due to the following three reasons: 1) the operator of $[\cdot]^+$ is non-smoothness for the objective function; 2) the variables in constraints \eqref{eq:st1}, \eqref{eq:st3}, and \eqref{eq:st4} are intricately coupling; 3) and they further are quadratic and fractional, which makes it more challenging to be solved. But Problem \eqref{eq:P1} can be simplified as follows: 1) the optimal solution of Problem \eqref{eq:P1} remains unchanged whether the $[\cdot]^+$ operator is included or omitted the operation, This condition inherently holds at the optimal solution because it automatically satisfies the non-negativity constraint enforced by the operator $[\cdot]^+$. For complete proof details, we refer the reader to Lemma 1 in \cite{Guangchi2019TWC}. 2) The fraction of $1/Q$ can also be removed as a constant and has no effects on the solution of Problem \eqref{eq:P1}, since scaling the objective function by a positive constant does not change the optimal solution. As a result, Problem \eqref{eq:P1} can be reduced to the following one.
\begin{subequations}\label{eq:P1*}
	\begin{align}
		\text{(P1*):} \max_{\mathbf{x}_\text{MA}[t], \mathbf{w}_{\text{MA}}[t], \mathbf{w}_{\text{FPA}}[t]}  &\sum_{t=1}^Q(R_{\rm bob}[t]-R_{\rm eve}[t]) \\
		\text{s.t.} \quad \quad   \text{Eq. }\eqref{eq:st1} &\text{ - Eq. } \eqref{eq:st4} . 
	\end{align} 
\end{subequations}

Although there is no general approach to solving Problem \eqref{eq:P1*} optimally, in the following section \ref{sec:optimization}, we propose an effective algorithm to find a high-quality feasible solution for Problem \eqref{eq:P1*}.

\section{Optimization Algorithm Design For FPA-MA Joint Design} \label{sec:optimization}
In this section, we present a joint optimization framework designed to maximize the secrecy rate between the Bob and colluding Eve. The algorithm simultaneously addresses three coupled parameters: 1) FPA's BF design, 2) MA's BF design, and 3) antenna positioning configuration of MA. To resolve this non-convex optimization challenge, we adopt an AO algorithm that decomposes the original problem into three tractable subproblems.
In detail, For BF optimization of FPA and MA, we reformulate their precoding tasks into Rayleigh quotient forms, subsequently deriving closed-form solutions through eigenvalue decomposition. The MA positioning optimization subproblem is addressed by a novel NMPGA algorithm that enhances search efficiency via parallel evolution mechanisms and adaptive mutation operators. These sub-solutions are iteratively optimized with our AO framework until convergence criteria are met, ultimately achieving a high-quality feasible solution with polynomial-time complexity.

\subsection{$\rm \mathbf{w}_{\rm FPA}[\it t]$ Optimization of {\rm FPA} with fixed $\mathbf{x}_{\rm MA}[\it t]$ and $\mathbf{w}_{\rm MA}[\it t]$} \label{subsection:P2}
For this optimization subproblem, positions $\mathbf{x}_\text{MA}[t]$ of the MA and BF vector $\mathbf{w}_{\text{MA}}[t]$ are fixed values. To simplify the symbolic expression and strip the fixed variables, the symbols of $\bm{\Gamma}_b$ and $\bm{\Gamma}_{e^*}$ are defined as
\begin{equation}
\begin{aligned}
	\bm{\Gamma}_b &= \frac{1}{\sigma^2 + G_\text{MA}(\theta_b)} \mathbf{a}(\mathbf{x}_{\text{FPA}}, \theta_b)\mathbf{a}^H(\mathbf{x}_{\text{FPA}}, \theta_b), \\
	\bm{\Gamma_{e^*}} &= \frac{1}{\sigma^2 + \sum_{i=1}^M G_\text{MA}(\theta_{e_{i}})}  \sum_{i=1}^M \mathbf{a}(\mathbf{x}_{\text{FPA}}, \theta_{e_{i}})\mathbf{a}^H(\mathbf{x}_{\text{FPA}}, \theta_{e_{i}}).
\end{aligned}
\end{equation}

Then, the problem (P1*) can be transformed into
\begin{subequations}\label{eq:P2}
	\begin{align}
		\text{(P2):} \quad & \max_{\mathbf{w}_{\text{FPA}}[t]} \frac{1+\mathbf{w}_{\text{FPA}}^H[t]\bm{\Gamma}_b\mathbf{w}_{\text{FPA}}[t]}{1+\mathbf{w}_{\text{FPA}}^H[t]\bm{\Gamma_{e^*}}\mathbf{w}_{\text{FPA}}[t]} \label{eq:8a} \\
		\text{s.t.} \quad & \text{Eq. } \eqref{eq:st4}. \label{eq:p2 c1}
	\end{align}
\end{subequations}

With Lemma 1 in \cite{Jia2024Physical}, the constraint \eqref{eq:p2 c1} can be transformed by the following theorem.

\theorem{The optimal BF vector $\mathbf{w}_{\text{FPA}}[t]$ of the problem  \eqref{eq:P2} must satisfy $||\mathbf{w}_{\text{FPA}}[t]||_2^2 = P_{\text{FPA}}$.}

\proof We prove this theorem by contradiction. Assuming that transmission power of the optimal solution is strictly less than the maximum available power, meaning $\|\mathbf{w}_{\text{FPA}}[t]\|_2^2 < P_{\text{FPA}}$. As a result, we can construct a new BF vector $\bar{\mathbf{w}}_{\text{FPA}}$, defined as $\bar{\mathbf{w}}_{\text{FPA}}[t] = \frac{\sqrt{P_{\text{FPA}}}}{\|\mathbf{w}_{\text{FPA}}[t]\|} \mathbf{w}_{\text{FPA}}[t]$, which satisfies $ \|\bar{\mathbf{w}}_{\text{FPA}}[t]\|_2^2 = P_{\text{FPA}} $. According to the non-decreasing nature of the objective function, $\bar{\mathbf{w}}_{\text{FPA}}[t]$ is a feasible solution and can yield a higher objective function value, meaning $R_{\rm sec}[t]^*(\bar{\mathbf{w}}_{\text{FPA}}[t])>R_{\rm sec}[t](\mathbf{w}_{\text{FPA}}[t])$, contradicting the assumption of optimality. Therefore, we must have $ \|\mathbf{w}_{\text{FPA}}[t]\|_2^2 = P_{\text{FPA}} $, completing the proof.

In order to transform The problem (P2) in Eq. \eqref{eq:P2} into a Rayleigh quotient solution problem, we introduce a normalized variable $\mathbf{z}$, $\mathbf{w}_{\text{FPA}}[t] = \sqrt{P_{\text{FPA}}}\mathbf{z}$ and $\|\mathbf{z}\|_2^2 = 1$. After replacing $\mathbf{w}_{\text{FPA}}[t]$ with $\mathbf{z}$, the original problem can be rewritten as
 \begin{subequations}
	\begin{align}
		\text{Objective:} \quad \max_{\mathbf{w}_{\text{FPA}}[t]} & \frac{1 + P_{\text{FPA}}\mathbf{z}^H\bm{\Gamma}_b\mathbf{z}}{1 + P_{\text{FPA}}\mathbf{z}^H\bm{\Gamma}_{e^*}\mathbf{z}} \quad  \\
		\text{s.t.} \quad &  \|\mathbf{z}\|_2^2 = 1 \label{eq:9b}.
	\end{align}
\end{subequations}

Then, the numerator and denominator are divided by $P_{\text{FPA}}$ at the same time, and the unit matrix transformation is introduced. The problem (P2) in Eq. \eqref{eq:P2} can be transformed into the Rayleigh quotient problem as follows
\begin{subequations}
	\begin{align}
		\text{Objective:} \quad \max_{\mathbf{w}_{\text{FPA}}[t]} & \frac{\mathbf{z}^H \left(\bm{\Gamma}_b + \frac{1}{P_{\text{FPA}}}\mathbf{I}_N\right)\mathbf{z}}{\mathbf{z}^H \left(\bm{\Gamma_{e^*}} + \frac{1}{P_{\text{FPA}}}\mathbf{I}_N\right)\mathbf{z}} \quad \label{eq:9a}\\
		\text{s.t.} \quad & \text{Eq. } \eqref{eq:9b}. 
	\end{align}
\end{subequations}
The optimal solution can be directly determined as \cite{2010TITKhisti}
\begin{equation}\label{eq:10}
\mathbf{w}^{*}_{\text{FPA}}[t] = \sqrt{P_{\text{FPA}}}\mathbf{V}_{\rm FPA}^{\max},
\end{equation}
where $\mathbf{V}_{\rm FPA}^{\max}$ is the normalized eigenvector corresponding to the largest eigenvalue of matrix $\left(\bm{\Gamma_{e^*}} + \frac{1}{P_{\text{FPA}}}\mathbf{I}_N\right)^{-1} \\ \left(\bm{\Gamma_}b + \frac{1}{P_{\text{FPA}}}\mathbf{I}_N\right)$.

\subsection{$\mathbf{x}_{\rm MA}[\it t]$ Optimization of {\rm MA} with fixed $\mathbf{w}_{\rm MA}[\it t]$ and $\mathbf{w}_{\rm FPA}[\it t]$}
For this optimization subproblem, the BF vectors of FPA and MA, namely  $\mathbf{w}_{\text{FPA}}[t]$ and $\mathbf{w}_{\text{MA}}[t]$, are fixed values. By introducing the Euler formula $e^{jy} = \cos(y) + j\sin(y)$, variables $\mathbf{c}_i$ and $\mathbf{s}_i$ are defined as
\begin{equation}\label{eq:12}
\begin{aligned}
	\mathbf{c}_i &= [c_{1,i}, \ldots, c_{n,i}, \ldots, c_{N,i}]^T, \quad i = 0,1,\ldots,M, \\
	\mathbf{s}_i &= [s_{1,i}, \ldots, s_{n,i}, \ldots, s_{N,i}]^T, \quad i = 0,1,\ldots,M,
\end{aligned}
\end{equation}
where the meanings of $c_{n,i}$ and $s_{n,i}$ are given by
\begin{equation}\label{eq:13}
c_{n,i} = \cos\left(\frac{2\pi}{\lambda}x_n[t] \cos \theta \right), \ 
s_{n,i} = \sin\left(\frac{2\pi}{\lambda}x_n[t] \cos \theta \right).
\end{equation}
In particular, the parameter $\theta$ is determined by the index $i$ based on the following condition:
\begin{displaymath}
\begin{aligned}
	\theta=
	\begin{cases}
		\theta_b,~~\text{if $i=0$}, \\
		\theta_{e_i},~~\text{otherwise},
	\end{cases}
\end{aligned}
\end{displaymath}
where $\theta_{e_i}$ denotes the parameter associated with the $i$-th entity for $i\neq0$.
Then, the following variables are defined: $\mathbf{w}_\text{MA} = \mathbf{u} + j\mathbf{z}$, $\mathbf{A} = \mathbf{u}\mathbf{u}^T + \mathbf{z}\mathbf{z}^T$, $\mathbf{S} = \mathbf{u}\mathbf{z}^T - \mathbf{z}\mathbf{u}^T$. As a result, the channel gain of MA to receivers can be represented as
\begin{equation}\label{eq:14}
\begin{aligned}
	|\mathbf{a}^H(\mathbf{x}_\text{MA}[t],\theta)\mathbf{w}_\text{MA}[t]|^2 = \mathbf{c}_i^T\mathbf{A}\mathbf{c}_i + \mathbf{s}_i^T\mathbf{A}\mathbf{s}_i + 2\mathbf{c}_i^T\mathbf{S}\mathbf{s}_i 
	\triangleq \gamma(\mathbf{c}_i,\mathbf{s}_i).
\end{aligned}
\end{equation}

The variable of $\gamma(\mathbf{c}_i,\mathbf{s}_i)$ represents the channel gain of the Bob with $i=0$, and the channel gain of the $i$-th Eve $e_i$ with $i\in [1,M]$. Consequently, the optimization problem (P1*) can be transformed into
\begin{subequations}\label{eq:P3}
	\begin{align} \notag
		\text{(P3):} \quad & \max_{\mathbf{x}_\text{MA}[t]} \log_2\left(1 + \frac{G_\text{FPA}(\theta_b)}{\gamma(\mathbf{c}_0, \mathbf{s}_0) + \sigma^2}\right)
 - \log_2\left(1 + \frac{\sum_{i=1}^M G_\text{FPA}(\theta_{e_i})}{\sum_{i=1}^M \gamma(\mathbf{c}_i, \mathbf{s}_i) + \sigma^2}\right) \\ 
		&\triangleq \Theta(\{\mathbf{c}_i, \mathbf{s}_i\}_{i=0}^M)\\
		\text{s.t.} \quad \quad \quad & \text{Eq. }\eqref{eq:st1} \text{ and Eq. } \eqref{eq:st2} . 
	\end{align}
\end{subequations}

The inherent non-convexity of optimization problem (P3), arising from coupled variables and intricate constraints, poses severe computational challenges. These difficulties are further compounded by the real-time operational constraints and stringent reliability requirements in multi-user scenarios, where the MA must undergo substantial positional adaptations across multiple sampling intervals. To reconcile these competing demands, we develop an improved PGA algorithm by integrating Nesterov momentum-based optimization principle \cite{nesterov1983method}, denoted by NMPGA. Compared to traditional momentum methods \cite{shiUnderstandingAccelerationPhenomenon2022}, Nesterov momentum incorporates a predictive gradient evaluation information at extrapolated positions prior to parameter updates, effectively mitigating optimization oscillations caused by eavesdropping collusion and antenna positioning projection constraints. Then, the updating rule for the antenna positioning vector can be represented as
\begin{subequations}\label{eq:16}
\begin{align}
	\mathbf{x}_\text{MA}^{tmp}[t] =& \mathbf{x}_\text{MA}^{i}[t] + \zeta \cdot v^{i}, \label{eq:16a} \\
	v^{i+1} = &\zeta \cdot v^i + \delta\nabla_{\mathbf{x}_\text{MA}^{tmp}[t]}\Theta(\{\mathbf{c}_i, \mathbf{s}_i\}_{i=0}^M), \label{eq:16b}\\
	\mathbf{x}_\text{MA}^{i+1}[t] =& \mathbf{x}_\text{MA}^{i}[t] + v^{i+1}, \label{eq:16c}\\
	\mathbf{x}_\text{MA}^{i+1}[t] =& \mathcal{P}\{\mathbf{x}_\text{MA}^{i+1}[t], d_{\text{min}}, L\} \label{eq:16d},
\end{align}
\end{subequations}
where $\zeta$ denotes the momentum coefficient, $v^i$ represents the velocity at the $i$-th iteration, and $\delta$ controls the gradient ascent step size. The optimization follows a sophisticated four-step iterative procedure. First, as shown in \eqref{eq:16a}, a temporary position $\mathbf{x}_\text{MA}^{tmp}[t]$ is updated through a momentum-directed advancement from the current position, enabling the algorithm to leverage historical information of antenna positions. The key innovation emerges in \eqref{eq:16b}, where the momentum vector update incorporates gradient information computed at this temporary position?a fundamental characteristic of Nesterov's method that achieves an optimal convergence rate of $\mathcal{O}(1/k^2)$, significantly surpassing the $\mathcal{O}(1/k)$ rate of standard gradient descent. This look-ahead gradient computation mechanism demonstrates particular efficacy in navigating challenging optimization landscapes, especially in regions characterized by narrow valleys or ill-conditioned geometries. The position update mechanism in \eqref{eq:16c} then synthesizes both historical momentum and predictive gradient information to facilitate an efficient exploration of the solution space. Finally, \eqref{eq:16d} applies the projection operator $\mathcal{P}\{\cdot\}$ to ensure the effectiveness of achieved solution within the constraints specified by \eqref{eq:st1} and \eqref{eq:st2}. Moreover, the operation of
$\mathcal{P}\{\cdot\}$ is defined as
\begin{equation}\label{eq:17}
\begin{aligned}
	& \mathcal{P}\{\mathbf{x}_\text{MA}^{i}[t], d_{\text{min}}, L\}: \\
	& \begin{cases}
		x_1^{i}[t] = \max(0, \min(L-(N-1)d_{\text{min}}, x_1^{i}[t])), \\
		x_2^{i}[t] = \max(x_1^{i}[t] + d_{\text{min}}, \min(L-(N-2)d_{\text{min}}, x_2^{i}[t])), \\
		\vdots \\
		x_r^{i}[t] = \max(x_{r-1}^{i}[t] + d_{\text{min}}, \min(L-(N-r)d_{\text{min}}, x_r^{i}[t])), \\
		\vdots \\
		x_N^{i}[t] = \max(x_{N-1}^{i}[t] + d_{\text{min}}, \min(L, x_N^{i}[t])).
	\end{cases}
\end{aligned}
\end{equation}

To further enhance optimization performance in complex solution spaces, we implement a dual adaptive mechanism that dynamically modulates both the momentum coefficient and step size parameters. This adaptive framework continuously adjusts these parameters based on optimization progress, increasing them to accelerate convergence during favorable conditions while implementing controlled reductions for refined local search when necessary. To sum up, the pesudo-code of NMPGA is given in \textbf{Algorithm \ref{alg:NMPGA}}.

\begin{algorithm}
	\small
	\caption{\small NMPGA algorithm}
	\label{alg:NMPGA}
	\begin{algorithmic}[1]
		\renewcommand{\algorithmicrequire}{\textbf{Input:}}
		\renewcommand{\algorithmicensure}{\textbf{Output:}}
		\REQUIRE maximum iterations $\mathbf{I}_{ter}$, convergence threshold $\tau_\mathbf{x}$, initial momentum $\zeta$, initial step size $\delta$ and velocity $v^0$
		\ENSURE $\mathbf{x}_{\text{MA}}^{\rm opt}[t]$
		\STATE Initialize: Store previous secrecy rate $R_{\rm sec}^{prev}$
		\REPEAT
		\STATE Compute $\mathbf{x}_\text{MA}^{tmp}[t]$ by Eq. \eqref{eq:16a}
		\STATE Update $v^{i+1}$ by Eq. \eqref{eq:16b}
		\STATE Update $\mathbf{x}_\text{MA}^{i+1}[t]$ by Eq. \eqref{eq:16c}
		\STATE Project $\mathbf{x}_\text{MA}^{i+1}[t]$ by Eq. \eqref{eq:16d}
		\STATE Calculate current secrecy rate $R_{\rm sec}^{curr}$
		\STATE Calculate trend by slide windows
		\IF{$R_{\rm sec}^{curr} > R_{\rm sec}^{prev}$}
		\STATE Increase $\delta$ and $\zeta$ slightly
		\STATE $\mathbf{w}_{\text{MA}}^\text{opt}[t] \Leftarrow \mathbf{w}_{\text{MA}}[t]$
		\ELSE
		\IF{$\text{trend} > 0$}
		\STATE Decrease $\delta$ and $v^i$ slightly
		\STATE Maintain $\zeta$ constant
		\ELSE
		\STATE Decrease $\delta$ and $\zeta$ slightly
		\STATE Decay $v^i$ strongly (not reset)
		\ENDIF
		\ENDIF
		\STATE $i \Leftarrow i + 1$
		\UNTIL{$|R_{\rm sec}^{curr} - R_{\rm sec}^{prev}| < \tau_\mathbf{x}$ \textbf{or} $i = \mathbf{I}_{ter}$}
	\end{algorithmic}
\end{algorithm}

In particular, the gradient of $\Theta(\{\mathbf{c}_i, \mathbf{s}_i\}_{i=0}^M)$ in Problem (P3) with respect to $\mathbf{x}_\text{MA}[t]$ can be derived as follows.
First, the gradient computation is initiated by evaluating the derivative of the exterior function $\gamma(\mathbf{c}_i,\mathbf{s}_i)$, resulting in the preliminary gradient formulation, as given in Eq.~\eqref{eq:grad_step_in} at the top of this page.
\begin{figure*}
\begin{equation}\label{eq:grad_step_in}
	\nabla_{\mathbf{x}_\text{MA}[t]}\Theta(\{\mathbf{c}_i, \mathbf{s}_i\}_{i=0}^M) =  \frac{1}{\ln 2} \left(\frac{-G_\text{FPA}(\theta_b)\nabla_{\mathbf{x}_\text{MA}[t]}\gamma(\mathbf{c}_0,\mathbf{s}_0)}{(\gamma(\mathbf{c}_0,\mathbf{s}_0) + \sigma^2)(1 + G_\text{FPA}(\theta_b))}
	- \frac{-\sum_{i=1}^M G_\text{FPA}(\theta_{e_i}) \sum_{i=1}^M \nabla_{\mathbf{x}_\text{MA}[t]}\gamma(\mathbf{c}_i,\mathbf{s}_i)}{(\sum_{i=1}^M\gamma(\mathbf{c}_i,\mathbf{s}_i) + \sigma^2)(1 + \sum_{i=1}^MG_\text{FPA}(\theta_{e_i}))}\right).
\end{equation}
\hrule 
\end{figure*}
Then, the gradient formula  for the interior function $\gamma(\mathbf{c}_i,\mathbf{s}_i)$ is
\begin{equation}\label{eq:20}
	\nabla_{\mathbf{x}_\text{MA}[t]} \gamma(\mathbf{c}_i,\mathbf{s}_i)= 
	\left[\frac{\partial \gamma(\mathbf{c}_i,\mathbf{s}_i)}{\partial x_1[t]}, \frac{\partial \gamma(\mathbf{c}_i,\mathbf{s}_i)}{\partial x_2[t]}, \ldots, \frac{\partial \gamma(\mathbf{c}_i,\mathbf{s}_i)}{\partial x_n[t]}, \ldots, \frac{\partial \gamma(\mathbf{c}_i,\mathbf{s}_i)}{\partial x_N[t]}\right], 
\end{equation}
where $ \in[0,1,2,\ldots,M]$. For each $n \in [1, 2, \ldots, N]$, to calculate the gradient corresponding to any antenna position $x_n[t]$ in a given time slot $t$, we introduce two intermediate functions $c_{n,i}$ and $s_{n,i}$ for conversion. That is, we can get
\begin{equation}\label{eq:21}
\begin{aligned}
	\frac{\partial \gamma(\mathbf{c}_i,\mathbf{s}_i)}{\partial x_n[t]} &= \frac{\partial \gamma(\mathbf{c}_i,\mathbf{s}_i)}{\partial c_{n,i}} \frac{\partial c_{n,i}}{\partial x_n[t]} + \frac{\partial \gamma(\mathbf{c}_i,\mathbf{s}_i)}{\partial s_{n,i}} \frac{\partial s_{n,i}}{\partial x_n[t]} \\
	&= -\frac{2\pi}{\lambda} \cos\theta \sin\left(\frac{2\pi}{\lambda} x_n[t] \cos\theta\right) \frac{\partial \gamma(\mathbf{c}_i,\mathbf{s}_i)}{\partial c_{n,i}} 
	+ \frac{2\pi}{\lambda} \cos\theta \cos\left(\frac{2\pi}{\lambda} x_n[t] \cos\theta\right) \frac{\partial \gamma(\mathbf{c}_i,\mathbf{s}_i)}{\partial s_{n,i}}.
\end{aligned}
\end{equation}

Furthermore, two diagonal matrix variables $\mathbf{D}_i$ and $\bm{\Lambda}_i$ are defined to represent the partial derivatives of $c_{n,i}$ and $s_{n,i}$ with respect to $x_n[t]$, respectively
\begin{equation}\label{eq:22}
\begin{aligned}
	\mathbf{D}_i &= \text{diag}\left\{\left[\frac{2\pi}{\lambda} \cos\theta \sin\left(\frac{2\pi}{\lambda} x_n[t] \cos\theta\right)\right]_{n=1}^N\right\}, \\[10pt]
	\bm{\Lambda}_i &= \text{diag}\left\{\left[\frac{2\pi}{\lambda} \cos\theta \cos\left(\frac{2\pi}{\lambda} x_n[t] \cos\theta\right)\right]_{n=1}^N\right\}.
\end{aligned}
\end{equation}

In addition, the partial derivatives of $\gamma(\mathbf{c}_i,\mathbf{s}_i)$ with respect to $c_{n,i}$ and $s_{n,i}$ can be calculated as
\begin{equation}\label{eq:23}
\begin{aligned}
	& \left[\frac{\partial \gamma(\mathbf{c}_i,\mathbf{s}_i)}{\partial c_{1,i}}, \frac{\partial \gamma(\mathbf{c}_i,\mathbf{s}_i)}{\partial c_{2,i}}, \ldots, \frac{\partial \gamma(\mathbf{c}_i,\mathbf{s}_i)}{\partial c_{n,i}}, \ldots, \frac{\partial \gamma(\mathbf{c}_i,\mathbf{s}_i)}{\partial c_{N,i}}\right]^T 
	 \triangleq \nabla_{\mathbf{c}_i} \gamma(\mathbf{c}_i,\mathbf{s}_i) = 2\mathbf{A}\mathbf{c}_i + 2\mathbf{S}\mathbf{s}_i, \\[10pt]
	& \left[\frac{\partial \gamma(\mathbf{c}_i,\mathbf{s}_i)}{\partial s_{1,i}}, \frac{\partial \gamma(\mathbf{c}_i,\mathbf{s}_i)}{\partial s_{2,i}}, \ldots, \frac{\partial \gamma(\mathbf{c}_i,\mathbf{s}_i)}{\partial s_{n,i}}, \ldots, \frac{\partial \gamma(\mathbf{c}_i,\mathbf{s}_i)}{\partial s_{N,i}}\right]^T 
	 \triangleq \nabla_{\mathbf{s}_i} \gamma(\mathbf{c}_i,\mathbf{s}_i) = 2\mathbf{A}\mathbf{s}_i - 2\mathbf{S}\mathbf{c}_i.
\end{aligned}
\end{equation}

Based on the above formulas, we can conclude that
\begin{equation}\label{eq:24}
\nabla_{\mathbf{x}_\text{MA}[t]} \gamma(\mathbf{c}_i,\mathbf{s}_i) = -\mathbf{D}_i(2\mathbf{A}\mathbf{c}_i + 2\mathbf{S}\mathbf{s}_i) + \bm{\Lambda}_i(2\mathbf{A}\mathbf{s}_i - 2\mathbf{S}\mathbf{c}_i).
\end{equation}

The final closed-form expression for gradient calculation is shown in Eq. \eqref{eq:grad} at the top of this page.
\begin{figure*}[t] 
\begin{equation} \label{eq:grad}
	\nabla_{\mathbf{x}_\text{MA}[t]}\Theta(\{\mathbf{c}_i, \mathbf{s}_i\}_{i=0}^M) =  \frac{1}{\ln 2} \left(\frac{\mathbf{D}_0(2\mathbf{A}\mathbf{c}_0 + 2\mathbf{S}\mathbf{s}_0) - \bm{\Lambda}_0(2\mathbf{A}\mathbf{s}_0 - 2\mathbf{S}\mathbf{c}_0)}{(\gamma(\mathbf{c}_0,\mathbf{s}_0) + \sigma^2)(1 + 1/G_\text{FPA}(\theta_b))}
	- \frac{\sum_{i=1}^M (\mathbf{D}_i(2\mathbf{A}\mathbf{c}_i + 2\mathbf{S}\mathbf{s}_i) - \bm{\Lambda}_i(2\mathbf{A}\mathbf{s}_i - 2\mathbf{S}\mathbf{c}_i)))}{(\sum_{i=1}^M\gamma(\mathbf{c}_i,\mathbf{s}_i) + \sigma^2)(1 + 1/\sum_{i=1}^MG_\text{FPA}(\theta_{e_i}))}\right).
\end{equation}
\hrule 
\end{figure*}
\subsection{$\mathbf{w}_\text{\rm MA}[t]$ Optimization of {\rm MA} with fixed $\mathbf{x}_\text{\rm MA}[t]$ and  $\mathbf{w}_{\text{\rm FPA}}[t]$}
For this optimization subproblem, the position $\mathbf{x}_\text{MA}[t]$ of the MA and BF vector $\mathbf{w}_{\text{FPA}}[t]$ of FPA are fixed values. Similar to the optimization method of dealing with BF design of FPA, to simplify the notation, the symbols of $\mathbf{K}_b$ and $\mathbf{K}_{e^*}$ are defined as follows
\begin{equation}\label{eq:26}
\begin{aligned}
	\mathbf{K}_b &= \mathbf{a}(\mathbf{x}_\text{MA}[t], \theta_b)\mathbf{a}^H(\mathbf{x}_\text{MA}[t], \theta_b), \\
	\mathbf{K}_{e^*} &= \sum_{i=1}^M \mathbf{a}(\mathbf{x}_\text{MA}[t], \theta_{e_i})\mathbf{a}^H(\mathbf{x}_\text{MA}[t], \theta_{e_i}).
\end{aligned}
\end{equation}

Then, the optimization problem (P1*) is transformed into
\begin{subequations}\label{eq:27}
\begin{align}
	\text{(P4):}~ & \max_{\mathbf{w}_{\text{MA}}[t]} \frac{1 + G_\text{FPA}(\theta_b)/(\sigma^2 + \mathbf{w}_{\text{MA}}^H[t]\mathbf{K}_b\mathbf{w}_{\text{MA}}[t])}{1 + \sum_{i=1}^M G_\text{FPA}(\theta_{e_i})/(\sigma^2 + \mathbf{w}_{\text{MA}}^H[t]\mathbf{K}_{e^*}\mathbf{w}_{\text{MA}}[t])} \\
	\text{s.t.} \quad & \text{Eq. }\eqref{eq:st3}.
\end{align}
\end{subequations}

Based on the derivation of transforming Problem (P2) into a Rayleigh quotient problem, as described in subsection \ref{subsection:P2}, a high-quality feasible solution can be obtained as \cite{2010TITKhisti}
\begin{equation}\label{eq:28}
\mathbf{w}^{*}_{\text{MA}}[t] = \sqrt{P_{\text{MA}}}\mathbf{V}_{\rm MA}^{\max}.
\end{equation}

Likewise, $\mathbf{V}_{\rm MA}^{\max}$ represents the normalized eigenvector corresponding to the largest eigenvalue of matrix $(\mathbf{I}_N + \frac{\sum_{i=1}^M G_\text{FPA}(\theta_{e_i}) \cdot \mathbf{I}_N}{\sigma^2 \cdot \mathbf{I}_N + P_{\text{MA}} \cdot \mathbf{K}_{e^*}})^{-1} (\mathbf{I}_N + \frac{G_\text{FPA}(\theta_b) \cdot \mathbf{I}_N}{\sigma^2 \cdot \mathbf{I}_N + P_{\text{MA}} \cdot \mathbf{K}_b})$.

\subsection{Alternating Optimization Algorithm}
Based on the derived analytical framework, the proposed AO algorithm with $\rm \mathbf{w}_{\rm FPA}[\it t]$, $\rm\mathbf{x}_{\text{\rm MA}}[\it t]$ and $\mathbf{w}_\text{\rm MA}[\it t]$ for determining the high-quality feasible solution at discrete time intervals proceeds through the following structured sequences. First, the process commences with the initialization of MA positions, followed by sequential optimization of three key parameters: the FPA's BF vector $\mathbf{w}_{\text{FPA}}[t]$, MA's BF vector $\mathbf{w}_{\text{MA}}[t]$, and MA spatial coordinates $\mathbf{x}_\text{MA}[t]$. During each parameter optimization phase, the remaining two variables are maintained as fixed constants to ensure convergence stability.
To overcome the inherent non-convexity of the optimization problem while satisfying the rigorous latency constraints in various wireless applications, we incorporate a multi-strategy optimization approach to enhance the efficiency of the AO algorithm $\rm \mathbf{w}_{\rm FPA}[\it t]$, $\rm\mathbf{x}_{\text{\rm MA}}[\it t]$ and $\mathbf{w}_\text{\rm MA}[\it t]$. This novel approach significantly improves solution quality and computational efficiency through two principal innovations: 1) We establish a dynamic benchmarking system that continuously updates the optimal secrecy rate during qualified iterations. This mechanism enables real-time tracking of global optima throughout the optimization trajectory, effectively addressing solution space variations; 2) A novel convergence criterion monitors successive iteration improvement rates via an early stopping strategy based on stagnation detection. This strategic implementation prevents premature convergence to local optimal solution, while maintaining computational efficiency-particularly crucial for high-mobility networks demanding ultra-low latency requirements. The pseudo-code of the proposed AO algorithm with $\rm \mathbf{w}_{\rm FPA}[\it t]$, $\rm\mathbf{x}_{\text{\rm MA}}[\it t]$ and $\mathbf{w}_\text{\rm MA}[\it t]$ is given in \textbf{Algorithm \ref{alg:AO}}.
\begin{algorithm}
	\small
	\caption{\small AO with $\mathbf{w}_{\text{FPA}}[t]$, $\mathbf{x}_{\text{MA}}[t]$ and $\mathbf{w}_{\text{MA}}[t]$}
	\label{alg:AO}
	\begin{algorithmic}[1]
		\renewcommand{\algorithmicrequire}{\textbf{Input:}}
		\renewcommand{\algorithmicensure}{\textbf{Output:}}
		\REQUIRE the default position vector $\mathbf{x}_\text{MA}^0[t]$, BF vector $\mathbf{w}_{\text{MA}}^0[t]$ of MA, BF vector $\mathbf{w}_{\text{FPA}}^0[t]$ of FPA, the maximum iteration number $\mathbf{I}_{ter}$, the convergence threshold $\tau_R$, initial index $i=0$, optimal secrecy rate $R_{\rm sec}^\text{opt}[t]=0$, stagnation counter $C_\text{stag} = 0$
		\ENSURE $\mathbf{w}_{\text{FPA}}^\text{opt}[t], \mathbf{w}_{\text{MA}}^\text{opt}[t] \text{ and } \mathbf{x}_\text{MA}^\text{opt}[t]$
		\REPEAT
		\STATE $i \Leftarrow i + 1$
		\STATE Update $\mathbf{w}_{\text{FPA}}^i[t]$ by Eq. \eqref{eq:10}
		\STATE Update $\mathbf{x}_\text{MA}^i[t]$ by NMPGA in Eq. \eqref{eq:16}
		\STATE Update $\mathbf{w}_{\text{MA}}^i[t]$ by Eq. \eqref{eq:27}
		\STATE $R_{\rm sec}^i[t] = R_{\rm sec}(\mathbf{x}_\text{MA}^i[t], \mathbf{w}_{\text{MA}}^i[t], \mathbf{w}_{\text{FPA}}^i[t])$
		\IF{$R_{\rm sec}^i[t] > R_{\rm sec}^\text{opt}[t]$}
		\STATE $\mathbf{w}_{\text{FPA}}^\text{opt}[t] \Leftarrow \mathbf{w}_{\text{FPA}}^i[t]$
		\STATE $\mathbf{w}_{\text{MA}}^\text{opt}[t] \Leftarrow \mathbf{w}_{\text{MA}}^i[t]$
		\STATE $\mathbf{x}_\text{MA}^\text{opt}[t] \Leftarrow \mathbf{x}_\text{MA}^i[t]$
		\STATE $R_{\rm sec}^\text{opt}[t] \Leftarrow R_{\rm sec}^i[t]$
		\STATE $C_\text{stag} \Leftarrow 0$
		\ELSE
		\STATE $C_\text{stag} \Leftarrow C_\text{stag} + 1$
		\IF{$C_\text{stag}$ reaches the stagnation threshold}
		\STATE \textbf{break}
		\ENDIF
		\ENDIF
		\UNTIL{$|R_{\rm sec}^i[t]-R_{\rm sec}^\text{opt}[t]| < \tau_R$ \textbf{or} $i = \mathbf{I}_{ter}$}
	\end{algorithmic}
\end{algorithm}

\textbf{\emph{Convergence:}} The optimization problem (P1), as defined in Eq. \eqref{eq:P1} and addressed by \textbf{Algorithm \ref{alg:AO}}, aims to maximize the average secrecy rate subject to the BF vectors of FPA and MA as well as MA's spatial coordinates. The average secrecy rate is monotonically non-decreasing due to the inherent nature of the optimization steps and is upper-bounded by the theoretical maximum secrecy rate, which is constrained by physical and system limitations. According to the monotone convergence theorem\cite{He2021Beamforming,Li2022Joint}, such a bounded monotonic sequence is guaranteed to converge. The algorithm's termination conditions, including the convergence threshold $\tau_R$ and maximum iteration number $\mathbf{I}_{ter}$, ensure a finite-time convergence. Furthermore, the stagnation detection implements controlled by early stopping strategy may halt optimization prior to reaching theoretical convergence thresholds, and confirms its essential convergence properties.

\textbf{\emph{Complexity:}} For each iteration of \textbf{Algorithm \ref{alg:AO}}, the optimization subproblems of FPA and MA's BF vector have the time complexity of $\mathcal{O}(N^3)$ \cite{Jia2024Physical}, mainly due to the decomposition of the eigenvalues. For the optimization of antenna positions, a NMPGA algorithm is employed. Gradient computation in each NMPGA iteration dominates the complexity, requiring $\mathcal{O}(M \cdot N^2)$ operations, where $M$ represents the number of Eves, as it involves matrix-vector multiplications and element-wise operations for each Eve. The projection of antenna positions onto the feasible set incurs a negligible complexity of $\mathcal{O}(N)$. With $\mathbf{I}^\text{M}_{ter}$ iterations in NMPGA, the total complexity for antenna position optimization is $O(\mathbf{I}^\text{M}_{ter} \cdot M \cdot N^2)$. Each iteration of AO comprises BF optimization for the FPA, BF optimization for the MA, and execution of NMPGA, leading to a per-iteration complexity of $\mathcal{O}(N^3 + \mathbf{I}^\text{M}_{ter} \cdot M \cdot N^2)$. Consequently, for $\mathbf{I}^\text{AO}_{ter}$ iterations of AO, the overall complexity of the \textbf{Algorithm \ref{alg:AO}} is $\mathcal{O}(\mathbf{I}^\text{AO}_{ter} \cdot (N^3 + \mathbf{I}^\text{M}_{ter} \cdot M \cdot N^2))$.

\section{Simulations and Performance Evaluations} \label{sec:numerical_results}
In this section, we rigorously evaluate the effectiveness of the proposed FMA co-design physical layer security framework. We conduct extensive simulation experiments using MATLAB simulator, implementing proposed NMPGA algorithm and comparing to conventional approaches.

\begin{figure}[htbp]
\centerline{\includegraphics[width=0.9\linewidth]{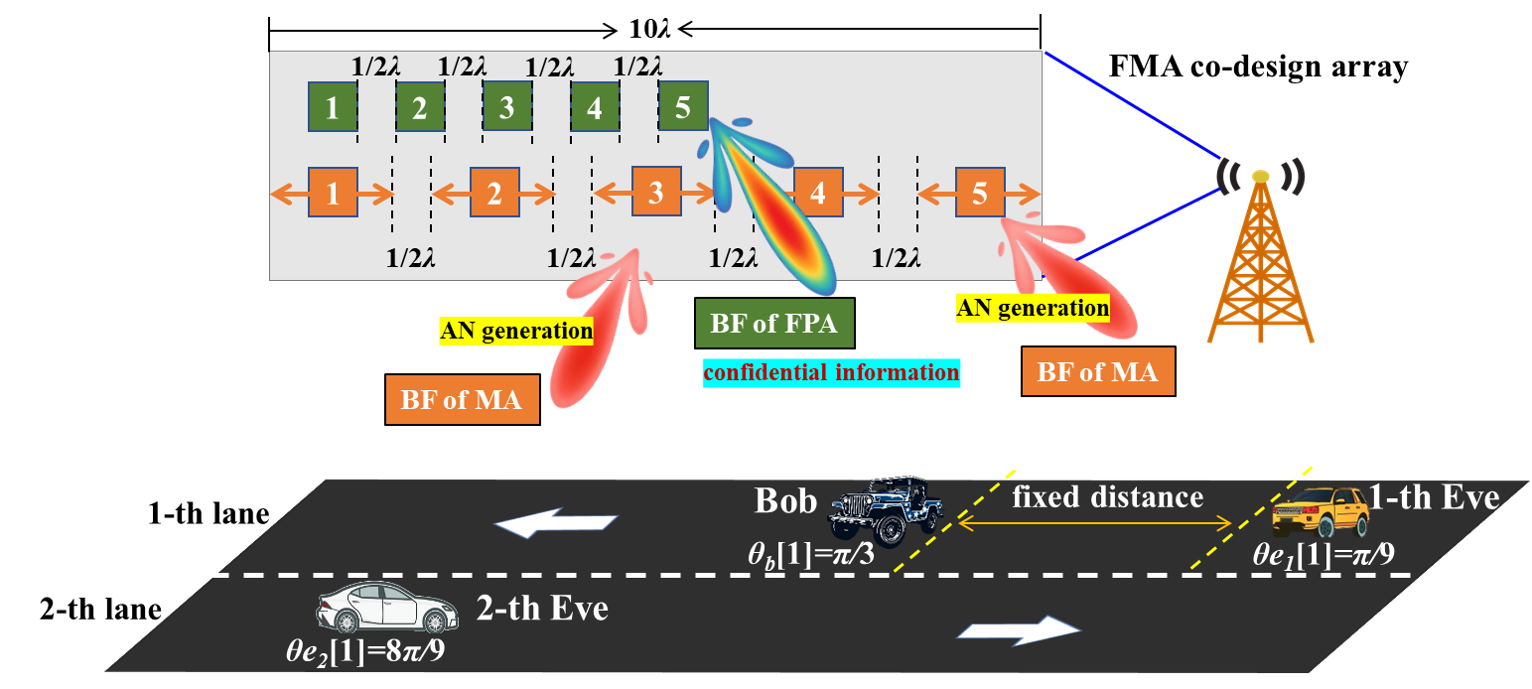}}
\caption{A simulation evaluation scenario}
\label{fig:scenario}
\end{figure}

\subsection{Simulation Settings}
Considering realistic vehicular communication paradigms, we establish a realistic vehicular communication scenario consisting of the BS equipped with hybrid FPA array and MA array, strategically deployed along a multilane highway, the Bob moving in the eastbound lane, while two adversarial Eves are intentionally positioned to represent distinct threat scenarios: one trailing Bob in the same lane with same velocity $v$ m/s and another approaching from the opposing lane with double velocity of Bob. All vehicles maintain uniform velocities to emulate real-world highway dynamics, ensuring continuous mobility patterns throughout the communication process.
The MA system is configured with a maximum displacement limit of $10\lambda$ and inter-element spacing constrained by $1/2\lambda$ to prevent electromagnetic coupling. To model the time-varying channel characteristics in dynamic environments, we discretize the angular domain through four time slots ($1\leq t\leq 4$) with the following azimuthal sampling:
\begin{displaymath}
	\theta_{b}[t] \in \{\frac{\pi}{3}, \frac{4\pi}{9}, \frac{5\pi}{9}, \frac{2\pi}{3}\},~ \theta_{e_1}[t] \in \{\frac{\pi}{9},~ \frac{2\pi}{9}, \frac{\pi}{3}, \frac{4\pi}{9}\},  \theta_{e_2}[t] \in \{\frac{8\pi}{9}, \frac{7\pi}{9}, \frac{2\pi}{3}, \frac{5\pi}{9}\},
\end{displaymath}
where the angular sets for the Bob, $1$-th Eve and $2$-th Eve emulate distinct relative moving trajectories. Key system parameters, including carrier frequency, transmission power, and noise power, are systematically cataloged in Table~\ref{tab:simulation_parameters}.

\begin{table}[h!]
	\small
\centering
\caption{\small Simulation Parameters and Settings}
\begin{tabular}{lp{6cm}l} 
	\toprule
	Parameter & Description & Value \\ 
	\midrule
	$P_{\text{FPA}}$ & transmission power of FPA & $5$W\textsuperscript{{\cite{Caban2019Design}}} \\
	$P_{\text{MA}}$ & transmission power of MA & $1$W\textsuperscript{\cite{Liu2025Covert}}\\
	$N$ & number of antennas & $5$\textsuperscript{\cite{Mei2024Movable-Antenna}} \\
	$M$ & number of Eves & $2$ \\
	$\lambda$ & wavelength & $0.0508$m \\
	$f_{c}$ & carrier frequency & $5.9$GHz\textsuperscript{{\cite{Zhang2024Sum-Rate}}} \\
	$\sigma^2$ & default noise power & $10^{-8}$W \\
	$d_{\textbf{min}}$ & min distance between antenna elements & $1/2\lambda$\textsuperscript{{\cite{Hu2024Movable-Antenna}}} \\
	$L$ & max effective movement position & $10\lambda$\textsuperscript{{\cite{Hu2024Secure}}} \\
	$\mathbf{I}_{ter}$ & maximum iteration number & $1500$ \\
	$\tau_\mathbf{x}$ & convergence threshold of $\mathbf{x}_\text{MA}$ & $10^{-6}$m \\
	$\tau_R$ & convergence threshold of $R_{\rm sec}$ & $10^{-6}$bps/Hz \\
	$\alpha$ &  path loss exponent & 2\textsuperscript{{\cite{hu2024reallyNeed}}} \\
	$d$ &  distance between BS and Bob or Eve & $100$m\textsuperscript{{\cite{hu2024reallyNeed}}}\\
	\bottomrule
\end{tabular}
\label{tab:simulation_parameters}
\end{table}

\subsection{Baseline Schemes}
To effectively evaluate the advantages of the proposed co-designing of FPA and MA (\textbf{FMA co-design}) enabled physical layer security framework, specifically the \textbf{FMA co-design} algorithm jointly optimizes the BF matrix of the FPA array and the BF matrix and antenna positions of the MA array, we establish two baseline configurations based on the traditional FPA and only MA arrays as follows.
\begin{itemize}
\item \textbf{MA-Only} \cite{Hu2024Secure, Xiong2025Analog}: All antennas are utilized to transmit the confidential information by jointly design the BF matrix and antenna positions of the BS. In particular, antenna position optimization is addressed by PGA algorithm with a fixed step size;
\item \textbf{FPA-Only} \cite{Zhu2016Improving,hu2019minimization}: The BF matrices for the confidential information and AN signals are optimized with 5 antennas, respectively. For the ease of representation, the \textbf{FPA-Only} baseline is classified into \textbf{FPA-Only-Conf. Infor.} and  \textbf{FPA-Only-AN}.
\end{itemize}

Moreover, in the studies of \cite{Lyu2025ISAC, Hu2024Secure} and \cite{ li2025MAEDF}, the PGA algorithm was utilized to address the sub-problem of antenna position optimization. However, due to the terminal mobility and low-latency communication requirements, this algorithm cannot be directly extended to optimize the antenna positions of the BS in our paper. To address this limitation, we proposed the NMPGA algorithm that can effectively handle the dynamic nature of our system while maintaining computational efficiency.
For further validating the convergence acceleration benefits of the proposed NMPGA algorithm when dealing with antenna positions optimization, this paper presents a comparative analysis between PGA and NMPGA algorithms under the \textbf{FMA co-design} framework, with different settings of the transmission power of the MA array, examining whether the approach can effectively address delay sensitivity issues while maintaining the overall system security.

\subsection{Convergence Assessment of the NMPGA and PGA Algorithm}
\begin{figure}[htbp]
	\centering
	\includegraphics[width=0.5\linewidth]{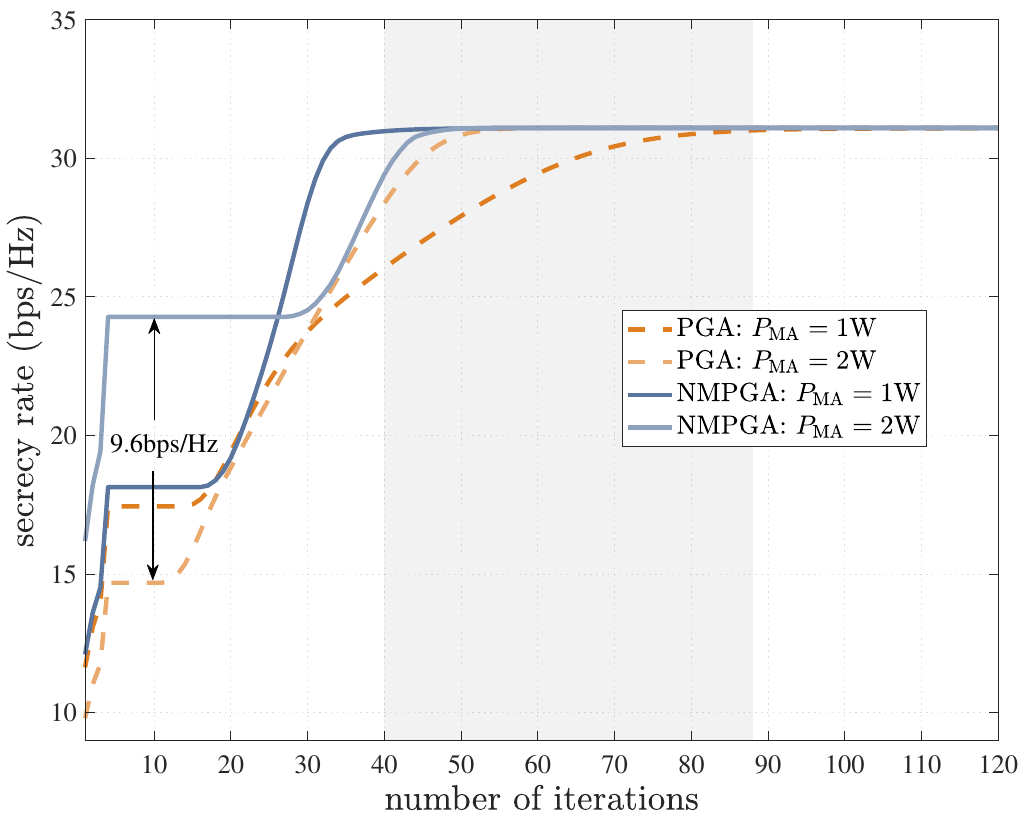}
	\caption{\small Secrecy rate vs. iterative}
	\label{fig:Rsec_iterations}
\end{figure}
The dynamic equilibrium state at $t=3$ is selected as the representative temporal snapshot for analysis in this subsection. This choice is motivated by two critical considerations. On one hand, this time slot captures critical spatial relationships where $\theta_b[3]=\frac{5\pi}{9}$, $\theta_{e_1}[3]=\frac{\pi}{3}$, and $\theta_{e_2}[3]=\frac{2\pi}{3}$.
On the other hand, this temporal snapshot is strategically selected due to its extreme parameter configuration among the sampled temporal instances. The observed dynamics at $t=3$ exhibit statistically significant alignment with neighboring time slots.  Crucially, convergence behavior at this critical snapshot serves as a rigorous performance lower bound, which demonstrates that if a rapid convergence is achievable under these most challenging conditions, the stability at less extreme temporal states is inherently guaranteed, confirming its validity for single-instance performance characterization.

Fig. \ref{fig:Rsec_iterations} demonstrates the superiority of the proposed NMPGA framework in terms of convergence dynamics and asymptotic performance. For the settings of $P_{\rm MA}=1$W, the NMPGA algorithm achieves a stable convergence ($\pm0.15$bps/Hz variance) at $R_s=31.1$bps/Hz within 40 iterations, exhibiting a distinct two-phase optimization pattern: 1) the average achievable secrecy rate can be increased from 12.1bps/Hz to 18.1bps/Hz in the 1-10-th iterations, and 2) further be enhanced from 18.1bps/Hz to 31.1bps/Hz in the range of [25, 40] iterations.
This contrasts sharply with the single-phase gradual convergence of the PGA algorithm \cite{Lyu2025ISAC, Hu2024Secure, li2025MAEDF}, requiring 88 iterations to reach value of $R_s=31$bps/Hz. The performance gap widens for the settings of $P_{\rm MA}=2$W, where NMPGA algorithm achieves two objectives: 
\begin{itemize}
\item the average achievable secrecy rate obtains an initial plateau of $R_s=24.3$bps/Hz within 5 iterations;
\item the rate converges to 30.9bps/Hz in the 45-th iteration,
\end{itemize} 
while the proposed PGA algorithm demonstrates a sub-optimal early-stage adaptation (achieving $R_s=14.7$bps/Hz) and delayed convergence (requiring 60 iterations) to achieve a stable performance at 31.1bps/Hz. As a comparison, the NMPGA algorithm outperforms \emph{early-stage gradient dominance and asymptotic stability margin}. In detail, it achieves a 64.6\% faster initial rate growth (4.8bps/Hz/iteration vs. 2.9bps/Hz/iteration), and keeps 0.0021bps/Hz advantage post-convergence.
To sum up, above analysis confirms the effectiveness and suitability of the NMPGA algorithm for security applications requiring a rapid convergence (less than 50 iterations) and millisecond-scale beam reconfiguration capabilities.

\subsection{Performance Analysis}
\begin{table}[htbp]
	\small
	\centering
	\caption{\small Antenna positions of MA array in 4 sample points}
	\begin{tabular}{cccccc} 
		\toprule
		Sample Point & $x_1[t]$ & $x_2[t]$ & $x_3[t]$ & $x_4[t]$ & $x_5[t]$ \\
		\midrule
		$\textbf{Initial}^*$ & $\mathbf{0.0000\lambda}$ & $\mathbf{0.5000\lambda}$ & $\mathbf{1.0000\lambda}$ & $\mathbf{1.5000\lambda}$ & $\mathbf{2.0000\lambda}$ \\
		t=1 & $\mathbf{0.0000\lambda}$ & $0.5001\lambda$ & $1.1349\lambda$ & $1.7677\lambda$ & $2.2678\lambda$ \\
		t=2 & $\mathbf{0.0000\lambda}$ & $\mathbf{0.5000\lambda}$ & $2.5959\lambda$ & $3.0959\lambda$ & $\mathbf{10.0000\lambda} \, \ $ \\
		t=3 & $0.2996\lambda$ & $0.7996\lambda$ & $2.4166\lambda$ & $3.0833\lambda$ & $9.7499\lambda$ \\
		t=4 & $0.2755\lambda$ & $1.0251\lambda$ & $2.3910\lambda$ & $2.8910\lambda$ & $9.9120\lambda$ \\
		\bottomrule
	\end{tabular}
	\label{tab:MA_position}
\end{table}
With the purpose of the average secrecy rate maximization, the antenna positioning optimization is addressed through the NMPGA algorithm. As observed in Table \ref{tab:MA_position}, the antenna positions undergo significant transformations during the optimization process. Initially, the MA array exhibits a uniform linear configuration with equidistant spacing of $0.5\lambda$ between adjacent elements. Through the iterative application of the NMPGA algorithm, \emph{this uniformity evolves into strategically non-uniform positioning to maximize the average secrecy rate}. Notably, certain antenna elements demonstrate interesting behavior throughout the optimization trajectory. For instance, the first antenna during the first two snapshot (i.e., $x_1[1] \text{ and } x_1[2]$) maintains its original position ($0.0000\lambda$) before shifting to non-zero positions in subsequent iterations. Similarly, the second antenna at temporal snapshot $t=2$ (i.e, $x_2[2]$) interestingly reverts exactly to its initial position ($0.5000\lambda$) after a slight deviation at $t=1$, suggesting that in this particular iteration during processing the optimization problem, the NMPGA algorithm determines that the original position for this specific antenna element coincidentally yielded better performance for the overall array configuration. Another significant observation is the behavior of the fifth antenna at temporal snapshot $t=2$ (i.e., $x_5[2]$), which reaches the maximum boundary constraint ($10.0000\lambda$), indicating that the gradient ascent optimization has driven this element to its positional limit to achieve optimal performance. These varied positional adaptations illustrate complex spatial dependencies in average secrecy rate optimization, and the effectiveness of the proposed NMPGA algorithm in navigating the multidimensional solution space to identify optimal antenna arrangements.
\begin{figure*}[htbp]
	\centering
	\subfloat[$t=1$]{\includegraphics[width=0.37\textwidth]{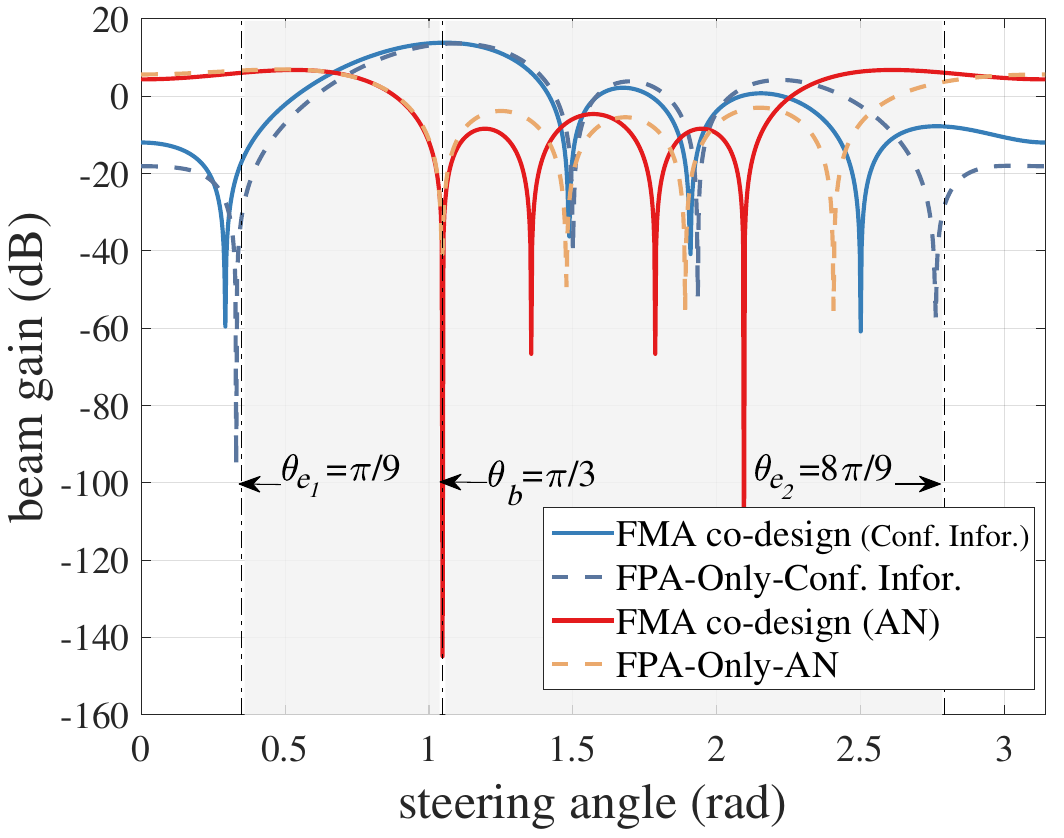}\label{fig:t=1}}
	\hfil
	\subfloat[$t=2$]{\includegraphics[width=0.37\textwidth]{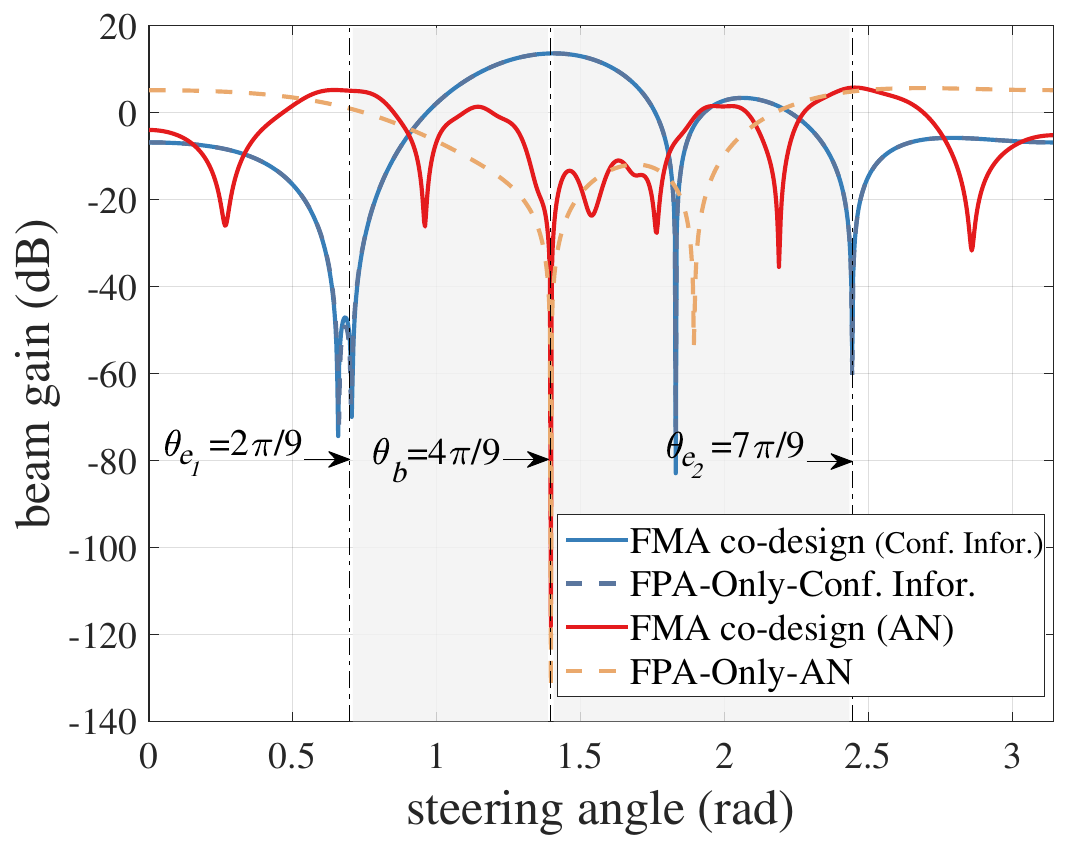}\label{fig:t=2}}
	
	\subfloat[$t=3$]{\includegraphics[width=0.37\textwidth]{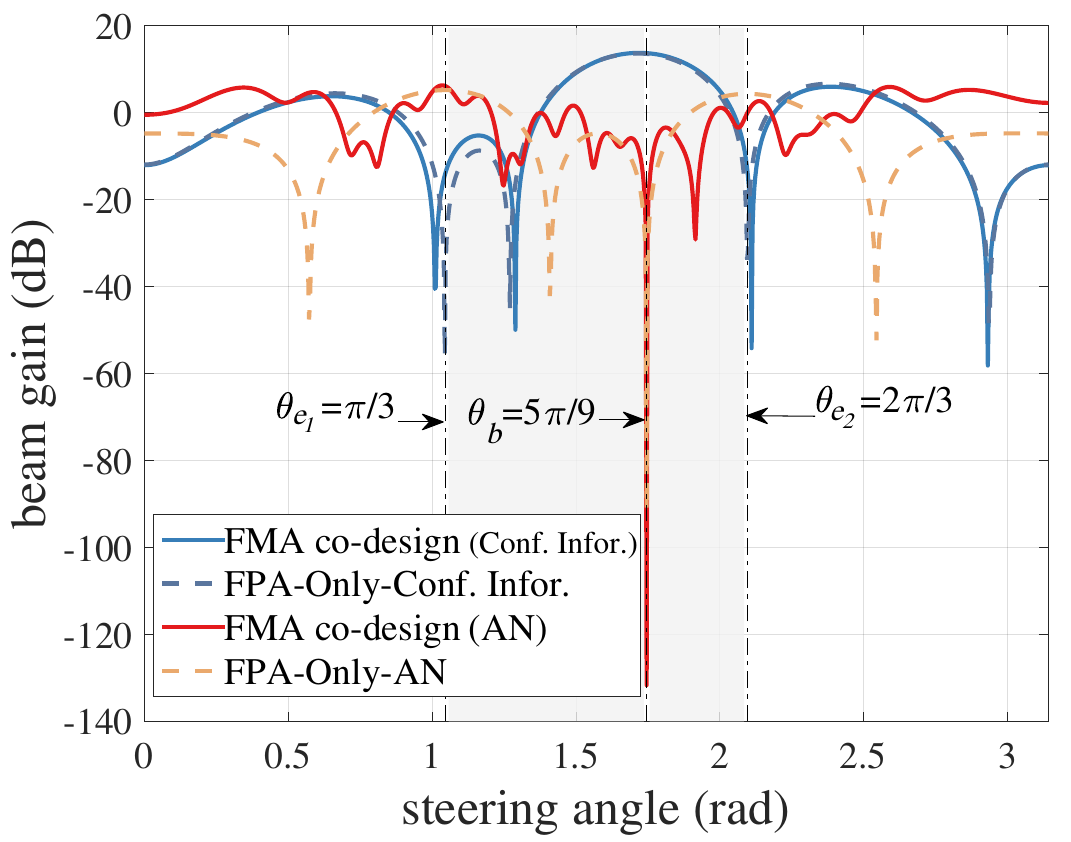}\label{fig:t=3}}
	\hfil
	\subfloat[$t=4$]{\includegraphics[width=0.37\textwidth]{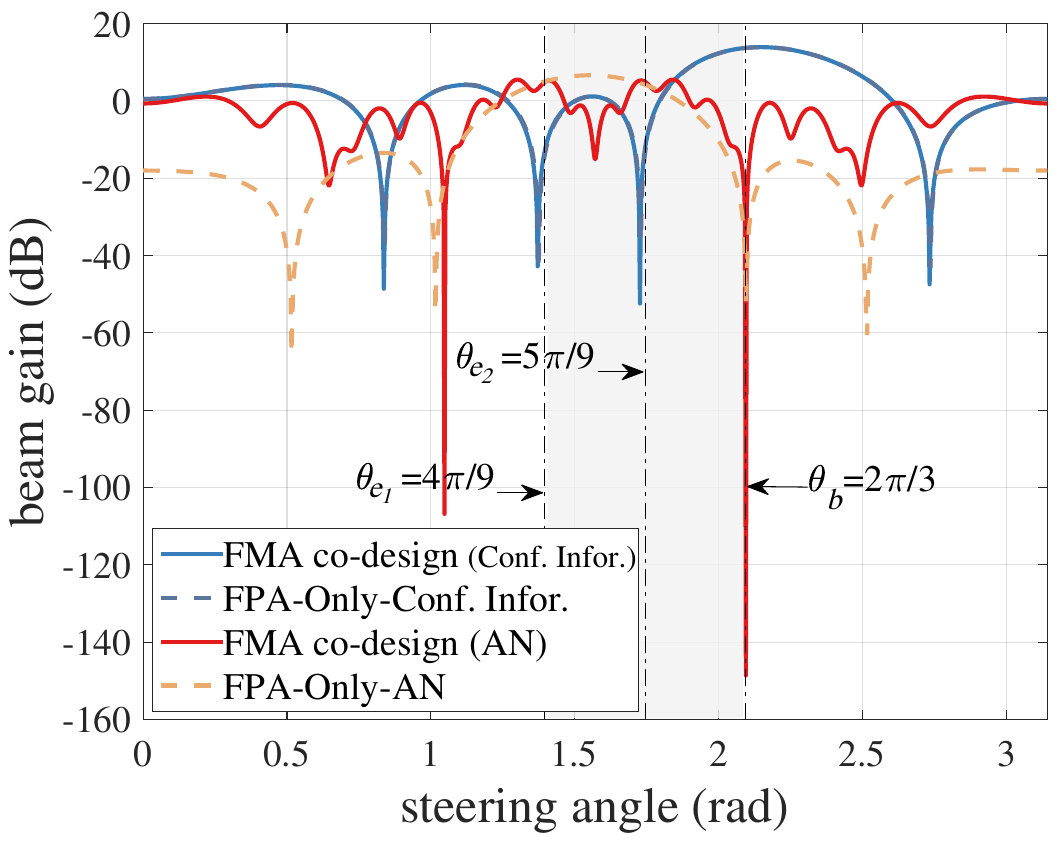}\label{fig:t=4}}
	
	\caption{\small Beam gain patterns at different time steps.}
	\label{fig:beam_patterns}
\end{figure*}

Based on five antenna positions in four time slots, as shown in Table \ref{tab:MA_position}, Fig. \ref{fig:t=1}-Fig. \ref{fig:t=4} depict the corresponding antenna gains received at the Bob and two Eves, respectively. In detail, for the Bob, the BF matrix design of the confidential information supported by the FPA array in proposed \textbf{FMA co-design} framework can achieve the most highest peak gains in all time slots, namely $13.44$dB, $13.60$dB, $13.50$dB, and $13.61$dB from the first to fourth time slots. Compared to \textbf{FPA-Only} \cite{Zhu2016Improving,hu2019minimization},  the difference of beam gains in all time slots can be ignored safely, since both beamforming approaches successfully make the maximum energy towards the Bob with similar efficiency, while creating significant nulls at the Eves' positions ($\theta_{e_1}$ and $\theta_{e_2}$). Notably, due to the selection of initial states for optimization and the algorithm terminating within an acceptable error margin, there are instances where the nulls in the confidential signal beam pattern may not perfectly align with the Eves' exact positions, especially in the 1st time slot. However, for secrecy rate maximization, further adjustments to the AN beam patterns effectively compensate for these slight misalignments by strategically directing interference power toward potential vulnerability regions.

Importantly, the crucial difference between above methods lies in the adaptive AN management capabilities. In other words, 
the proposed \textbf{FMA co-design} framework can achieve a superior interference suppression accuracy. For example, in the first time slot, as shown in Fig. \ref{fig:t=1}, the generated AN beam gains to the 1-th Eve ($\theta_{e_1}=0.35$rad) and 2-th Eve ($\theta_{e_2}=2.79$rad) are $6.13$dB and $6.07$dB, while only suppressing interference to the Bob ($\theta_{b}=1.05$rad) with $-144.89$dB. 
In contrast, the \textbf{FPA-Only} baseline exhibits uncoordinated interference management,  with AN beam gains demonstrating significant fluctuations across adversarial and legitimate channel. Namely $6.70$dB and $3.73$dB at the 1-th Eve ($\theta_{e_1}=0.35$rad) and 2-th Eve ($\theta_{e_2}=2.79$rad), as well as a $-41.24$dB at the Bob ($\theta_{b}=1.05$rad),
indicating an insufficient spatial selectivity and potential security leakage. This instability stems from FPA's fixed steering vector constraints, which limit its ability to decouple AN radiation patterns from legitimate signal paths. Compared to the proposed \textbf{FMA co-design} framework, the AN-BF supported by \textbf{FPA-Only-AN} baseline suffers from 1.77dB reductions totally in Eve-targeted interference efficiency, and a 103.65dB interference improvement at the Bob ($\theta_{b}=1.05$rad), directly impacting average achievable secrecy rates. The \textbf{FMA co-design} framework demonstrates a 18.6dB superior interference suppression relative to \textbf{FPA-Only-AN} baseline \cite{Zhu2016Improving,hu2019minimization}.

The beamforming architectures exhibit consistent temporal adaptability across following sequential time slots ($t={2,3,4}$) through dynamic reconfiguration, maintaining pattern similarity with the baseline of $t=1$. For example, from $t=1$ to $t=2$, the \textbf{FMA co-design} framework exhibits effectively the seamless beam redirection of the confidential information to the Bob from $\theta_b=\frac{\pi}{3}$ to $\theta_b=\frac{4\pi}{9}$, while simultaneously reshaping AN patterns synchronously with Eves' mobility, namely $\theta_{e_1}=\frac{\pi}{9}$ to $\theta_{e_1}=\frac{2\pi}{9}$ for the 1-the Eve and $\theta_{e_2}=\frac{8\pi}{9}$ to $\theta_{e_2}=\frac{7\pi}{9}$ for the 2-the Eve. In particular, on one hand, in the 1-th time slot, the \textbf{FMA co-design (Conf. Infor.)} framework demonstrates a 0.33dB superior beam gain of the confidential information relative to the \textbf{FPA-Only-Conf. Infor.} baseline \cite{Zhu2016Improving,hu2019minimization}. On the other hand, the \textbf{FMA co-design (AN)} framework reveals a consistent superiority of interference suppression across all temporal snapshots at the Bob (below -100dB), regardless of angular variations. For example, in the 4-th time slot, the \textbf{FMA co-design (AN)} framework demonstrates a 96.51dB superior interference suppression relative to \textbf{FPA-Only-AN} baseline \cite{Zhu2016Improving,hu2019minimization}. Featured by flexible beamforming management of the confidential information and AN signals, the \textbf{FMA co-design} framework sustains substantial performance advantages in dynamic environments, enhancing reliability for physical layer security implementations.

\begin{figure*}[htbp]
	\centering
	\subfloat[$t=1$]{\includegraphics[width=0.37\textwidth]{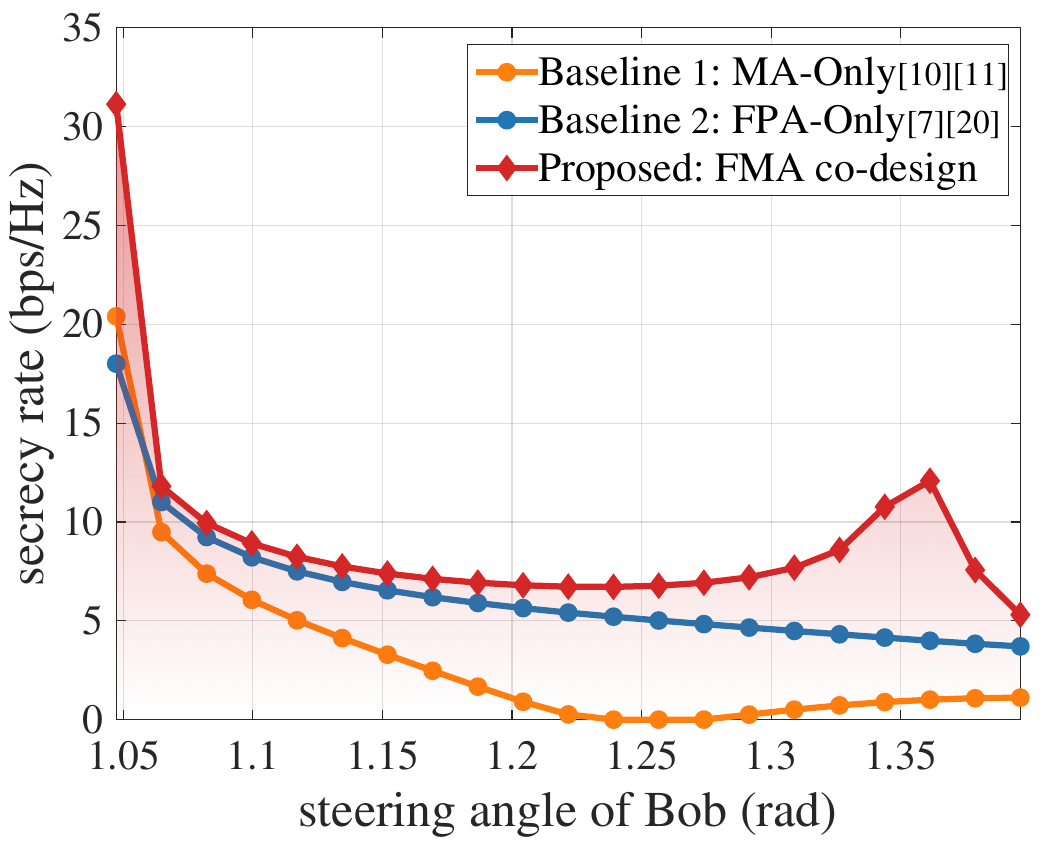}\label{fig:t=1r}}
	\hfil
	\subfloat[$t=2$]{\includegraphics[width=0.37\textwidth]{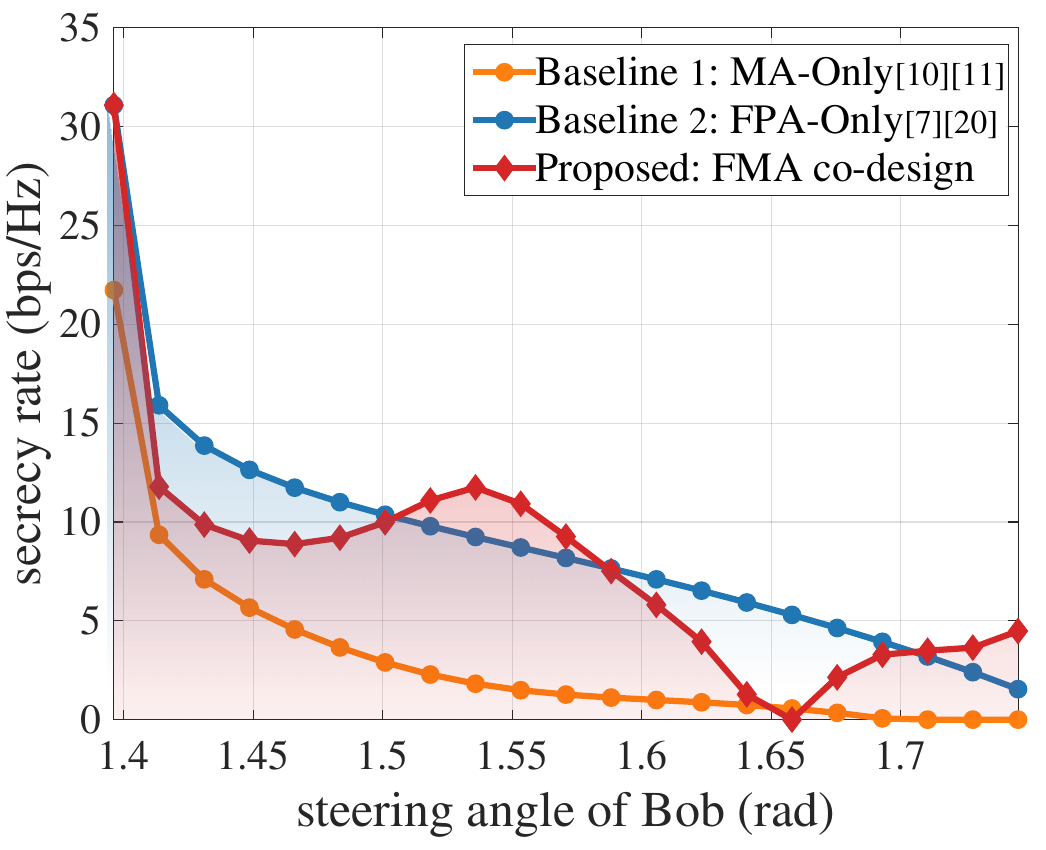}\label{fig:t=2r}}
	
	\subfloat[$t=3$]{\includegraphics[width=0.37\textwidth]{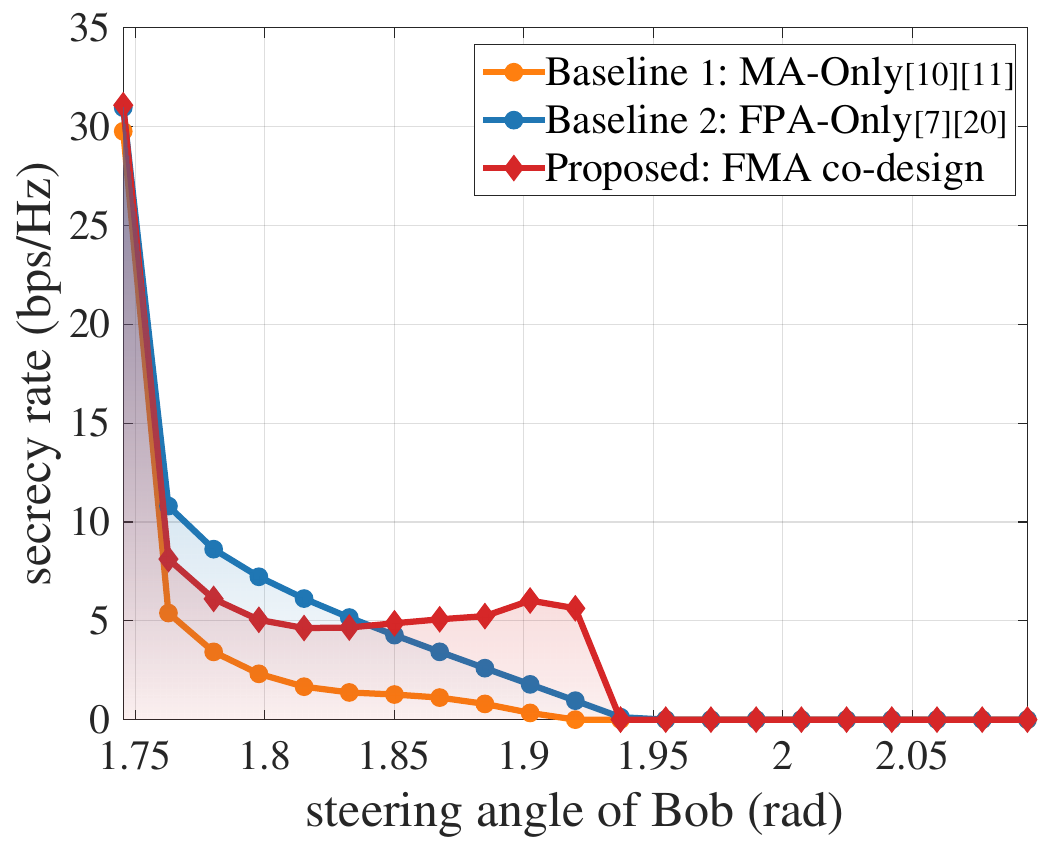}\label{fig:t=3r}}
	\hfil
	\subfloat[$t=4$]{\includegraphics[width=0.37\textwidth]{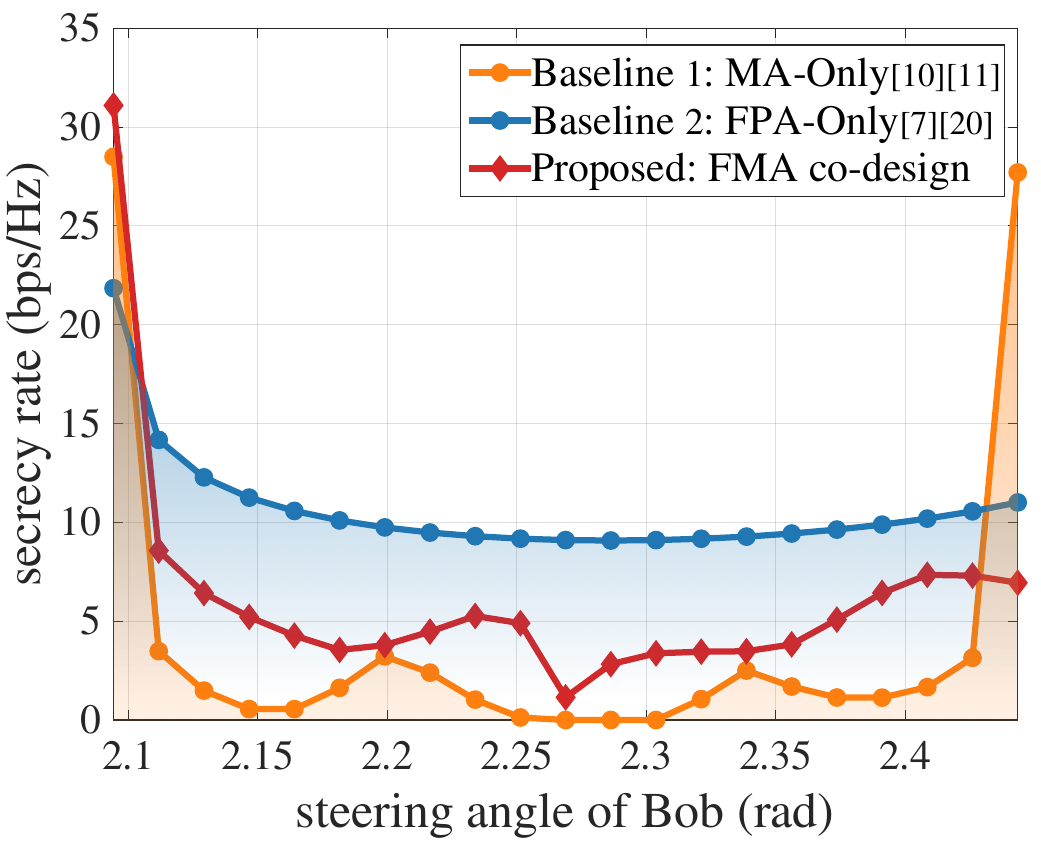}\label{fig:t=4r}}
	
	\caption{\small Secrecy rate vs. the steering angle of Bob.}
	\label{fig:robustness}
\end{figure*}

In four time slots, Fig.~\ref{fig:robustness} evaluates the secrecy rate for three schemes under different moving trajectories for the Bob, 1-th Eve, and 2-th Eve. In the $1$-th time slot, as shown in Fig. \ref{fig:t=1r}, the \textbf{FMA co-design} framework consistently achieves superior initial secrecy rates of 31.1bps/Hz at least across all four time slots, demonstrating the effectiveness of \textbf{Algorithm \ref{alg:AO}}). For comparison, on one hand, the \textbf{FPA-Only} baseline achieves an initial secrecy rate of $18.0$bps/Hz, making a $42.1\%$ lower than the proposed framework, primarily due to its limited spatial DoF in interference management. On the other hand, the \textbf{MA-Only} baseline \cite{Hu2024Secure, Xiong2025Analog}, while offering spatial flexibility, achieves initial rates of $20.4$bps/Hz, revealing the limitations of relying solely on the MA array for ensuring the performance of communication and security. Similar phenomenons exist in the subsequent time slots. In the 2-th time slot, although the \textbf{FMA co-design} framework generally yields to the \textbf{FPA-Only} approach, it still maintains localized performance superiority within specific angular regions (such as from $\theta_b=1.50$ to $1.58$rad and from $\theta_b=1.84$ to $1.93$rad),  as described in Fig. \ref{fig:t=2r}. In contrast, the \textbf{MA-Only} baseline consistently underperforms relative to the \textbf{FMA co-design} framework across all temporal snapshots.

Furthermore, the \textbf{MA-Only} baseline \cite{Hu2024Secure, Xiong2025Analog} exhibits the most rapid performance deterioration despite its theoretically superior DoF. In contrast, the \textbf{FPA-Only} baseline\cite{Zhu2016Improving,hu2019minimization} manifests a relatively stable degradation pattern with gradual performance decay over different time slots. The proposed \textbf{FMA co-design} framework demonstrates a particularly notable phenomenon through its distinctive non-monotonic trajectory. The hybrid architecture initially experiences performance degradation, followed by a partial recovery phase before entering subsequent decline cycles, ultimately establishing multiple local maxima (located at $\theta_b=1.36$, $1.54$, $1.90$, and $2.23$rad) interspersed with transient minima. This oscillatory pattern fundamentally explains its superior performance relative to the \textbf{FPA-Only} baseline \cite{Zhu2016Improving,hu2019minimization} within specific angular intervals, while simultaneously revealing its operational limitations in non-optimal angular regions.

Finally, because of the extreme parameters configuration at snapshot $t=3$, all methods (including the proposed \textbf{FMA co-design} framework) fail to achieve a positive when the positions of the Bob exceeds $\theta_b=1.94$rad, highlighting the fundamental challenges in maintaining secure communication under certain critical geometric configurations of the Bob and Eves.

\begin{figure}[htbp]
	\centering
	\includegraphics[width=0.6\linewidth]{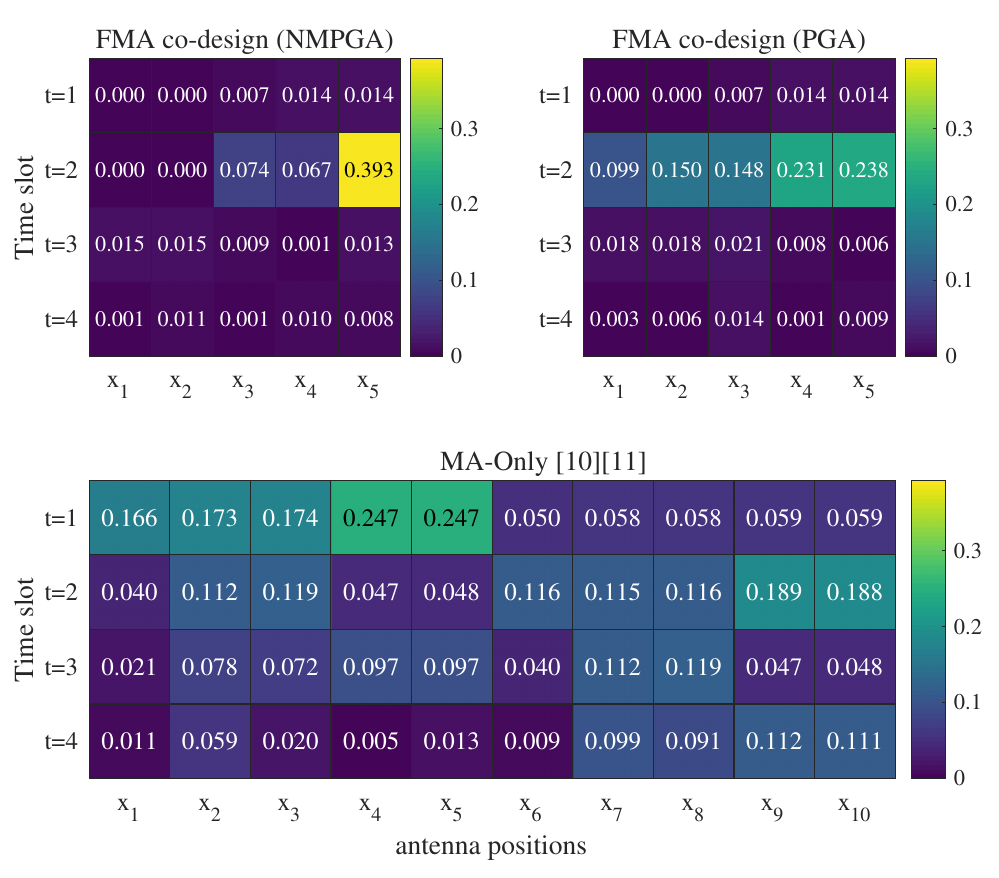}
	\caption{\small Antenna movement in different time slot}
	\label{fig:movement}
\end{figure}

\begin{figure*}[htbp]
	\centering
	\subfloat[Channel capacity $C_{b}$ vs. noise power]{\includegraphics[width=2in]{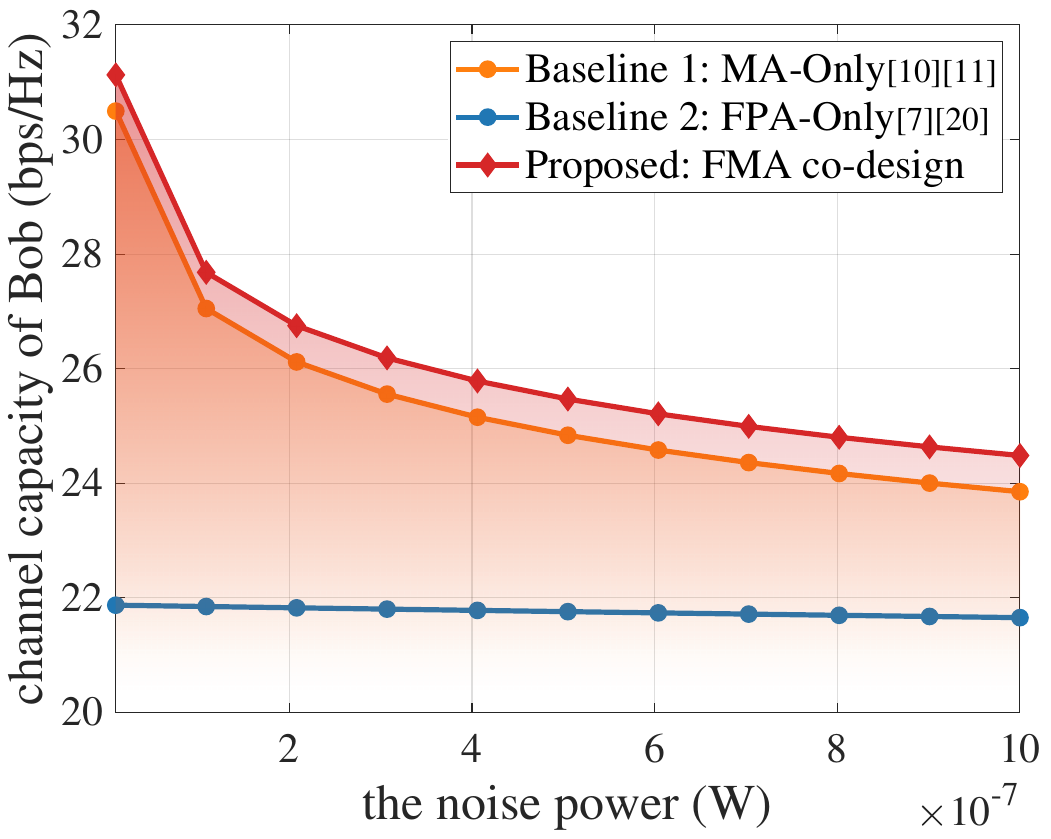}\label{fig:noise_a}}
	\subfloat[Channel capacity $C_{e^*}$ vs. noise power]{\includegraphics[width=2.02in]{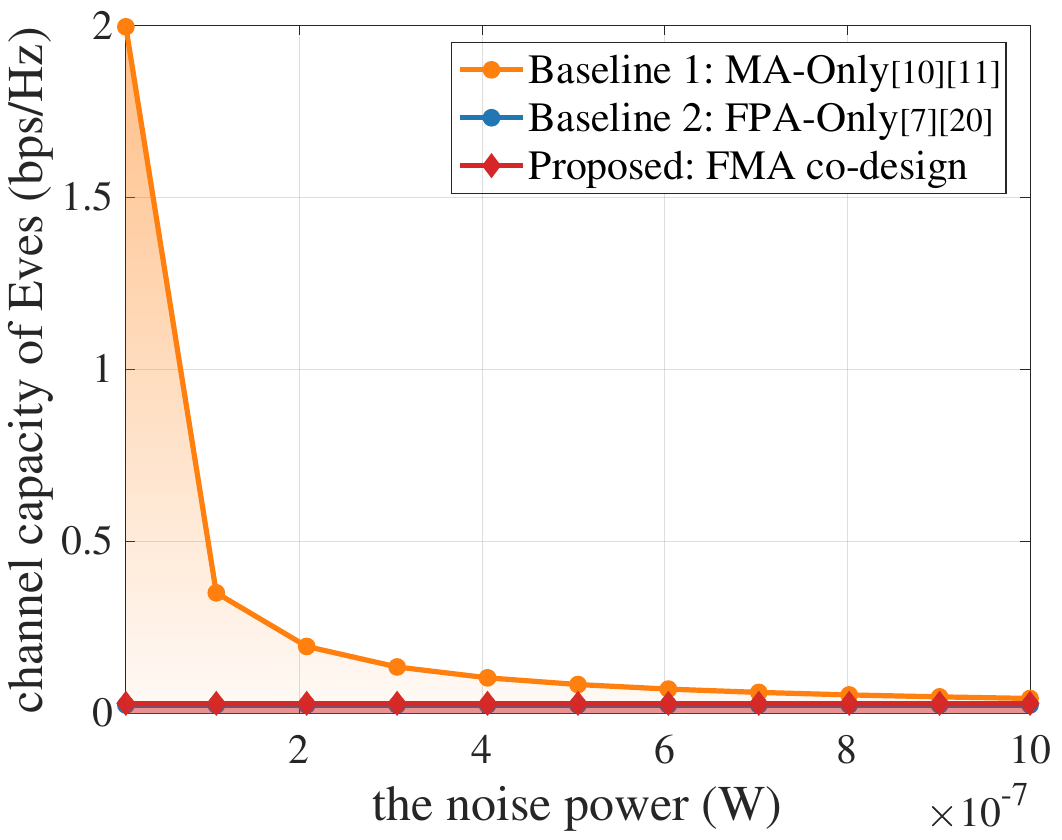}\label{fig:noise_b}}
	\subfloat[Secrecy rate vs. noise power]{\includegraphics[width=2in]{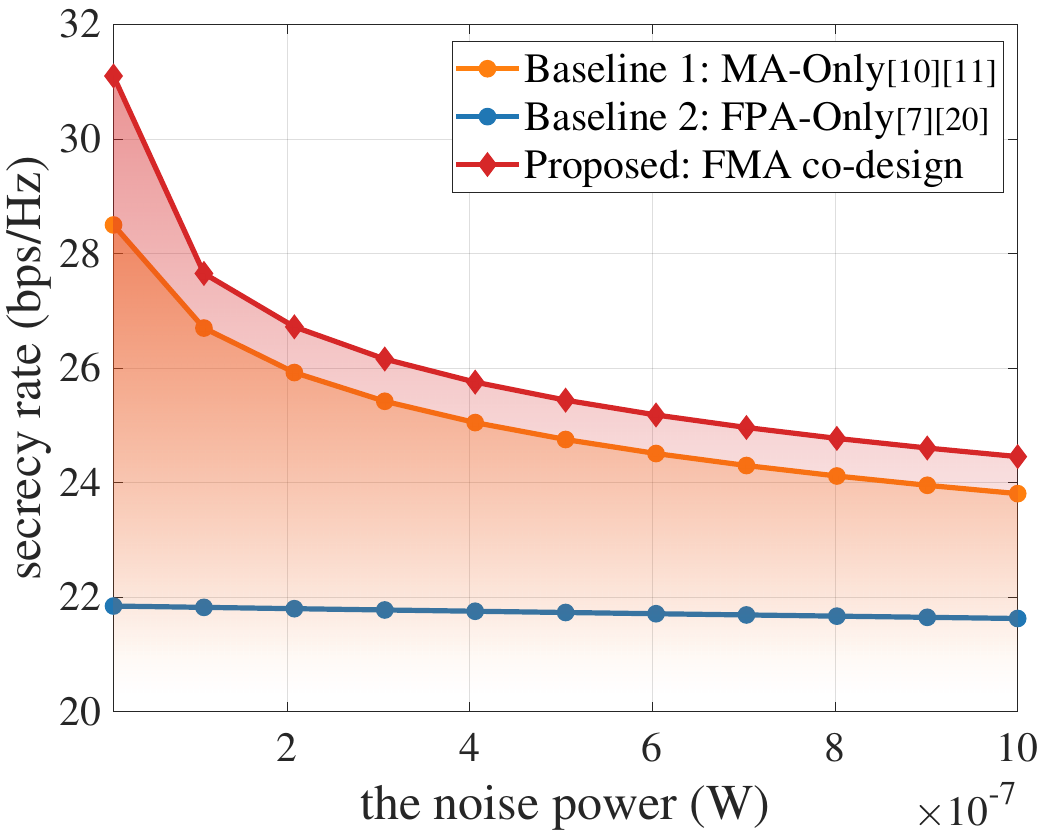}\label{fig:noise_c}}
	
	\caption{\small Channel capacity and secrecy rate for different settings of noise power.}
	\label{fig:noise_power}
\end{figure*}

Fig. \ref{fig:movement} quantifies the spatial displacement of antennas across successive time slots for the \textbf{FMA co-design} and \textbf{MA-Only} methods. In the \textbf{FMA co-design} framework, NMPGA algorithm achieves focused positional adjustments (the maximum displacement: 0.393m for the 5-th antenna ($x_5$) at $t=2$), whereas PGA algorithm adopts a balanced redistribution strategy (displacement range: from 0.099m to 0.238m for the 5-th antenna ($x_5$) at $t=2$). Critically, both algorithms suppress antenna movement during the 3-th time slot and 4-th time slot, particularly for the 5-th antenna ($x_5$), two algorithms achieve 0.383m and 0.126m reductions in displacement range compared to these in the 2-th time slot, which underscores their energy-efficient operation.

The \textbf{MA-Only} baseline \cite{Hu2024Secure, Xiong2025Analog} deploys ten antennas ($x_1\rightarrow x_{10}$) with dual transmission power allocation: $x_1\rightarrow x_5$ with 5W and $x_6\rightarrow x_{10}$ with 1W. The former exhibits clustered mobility patterns, peaking at $x_4$ and $x_5$ (0.247m at $t=1$), while the latter demonstrates a decentralized motion with maxima at $x_9$ and $x_{10}$ (0.189m and 0.188m at $t=2$). This bifurcated mobility distribution reflects inherent trade-offs in MA systems-contrary to the proposed \textbf{FMA co-design} framework, which strategically integrates static arrays for coverage stability and dynamically repositions a subset of antennas to maximize secrecy performance.

Fig. \ref{fig:noise_power} systematically evaluates the noise resilience of communication security through three key metrics: Bob's channel capacity, Eve's channel capacity, and secrecy rate. It can be noticed from Fig. \ref{fig:noise_a} that the \textbf{FMA co-design} framework obtains a 31.1bps/Hz initially and decreases to 24.5bps/Hz at the maximum noise power. At the critical noise level of  $10 \times 10^{-7}$W, the proposed framework sustains a 24.5 bps/Hz legitimate channel capacity-0.6bps/Hz and 2.8bps/Hz superior to \textbf{MA-Only} and \textbf{FPA-Only} baselines, respectively, demonstrating robust consistent security preservation. 
For Eve's channel capacity shown in Fig. \ref{fig:noise_b}, the \textbf{FMA co-design} framework keeps an adversarial channel capacity to 0.028 bps/Hz ($\pm5.7\%$ variance) across all noise conditions. While the \textbf{MA-Only} baseline initially permits 2.0bps/Hz at the minimum noise power, rapidly decreasing to 0.3bps/Hz at the maximum noise power, exposing fundamental instability. The \textbf{FPA-Only} baseline though marginally better in security (0.023bps/Hz), sacrifices 34.2\% legitimate channel capacity relative to the \textbf{FMA co-design} framework.

Next, as shown in Fig. \ref{fig:noise_c}, the \textbf{FMA co-design} framework exhibits exceptional noise tolerance, maintaining a 78.8\% secrecy rate retention ($31.1\rightarrow24.5$bps/Hz) as the noise power escalates from $1 \times 10^{-7}$W to $10 \times 10^{-7}$W. This outperforms the \textbf{MA-Only} baseline, which suffers a 16.5\% larger degradation ($28.5\rightarrow23.8$ bps/Hz),  the \textbf{FPA-Only} baseline that stagnates at 21.8 bps/Hz ($\pm0.5\%$ variation), highlighting \textbf{FMA co-design} framework's adaptive noise compensation capability.

These results conclusively demonstrate the \textbf{FMA co-design} framework's dual-capability paradigm: 1) Maintaining a 78\% of peak secrecy rate at least under $10\times$ noise escalation, and 2) Enforcing sub-0.03 bps/Hz eavesdropping channel capacity regardless of noise fluctuations. The architecture strategically combines fixed antennas' stability with movable elements' reconfigurability, achieving a 1.4-3.2$\times$ improvement in noise resilience over conventional \textbf{FPA-Only} and \textbf{MA-Only} methods.
\begin{figure*}[!t]
	\centering
	\subfloat[Channel capacity $C_{b}$ vs. path-loss exponent]{\includegraphics[width=2in]{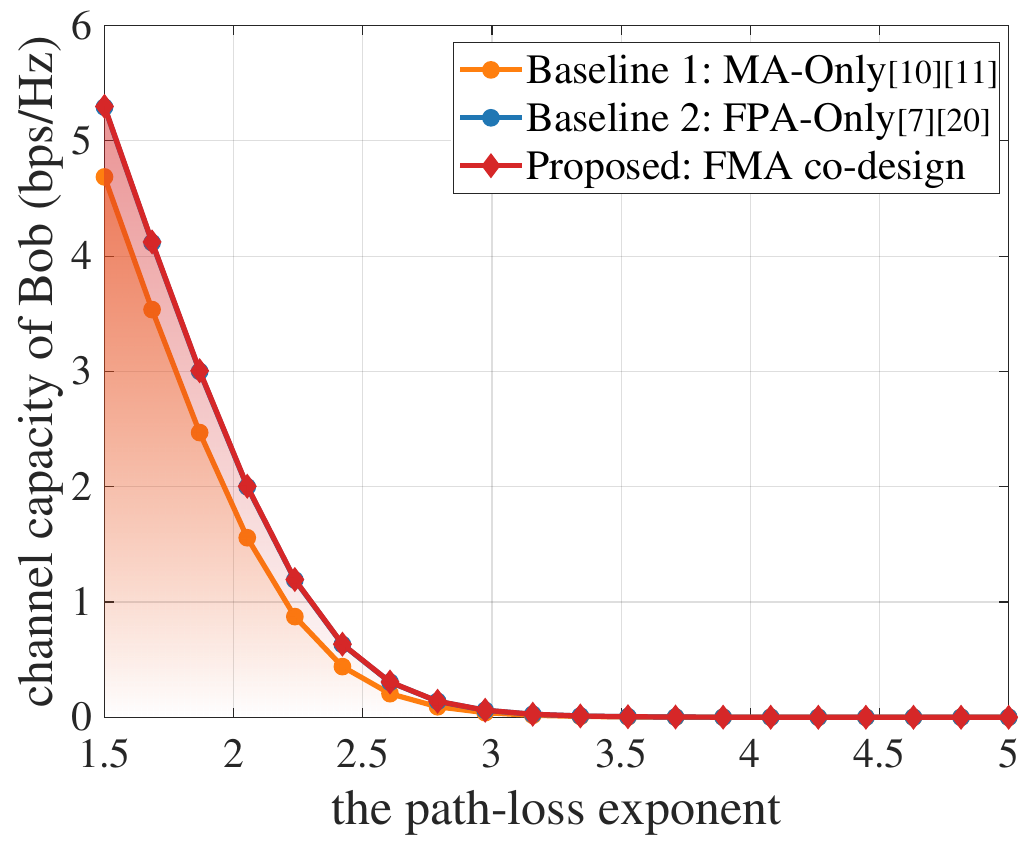}\label{fig:path_loss_a}}
	\subfloat[Channel capacity $C_{e^*}$ vs. path-loss exponent]{\includegraphics[width=2.1in]{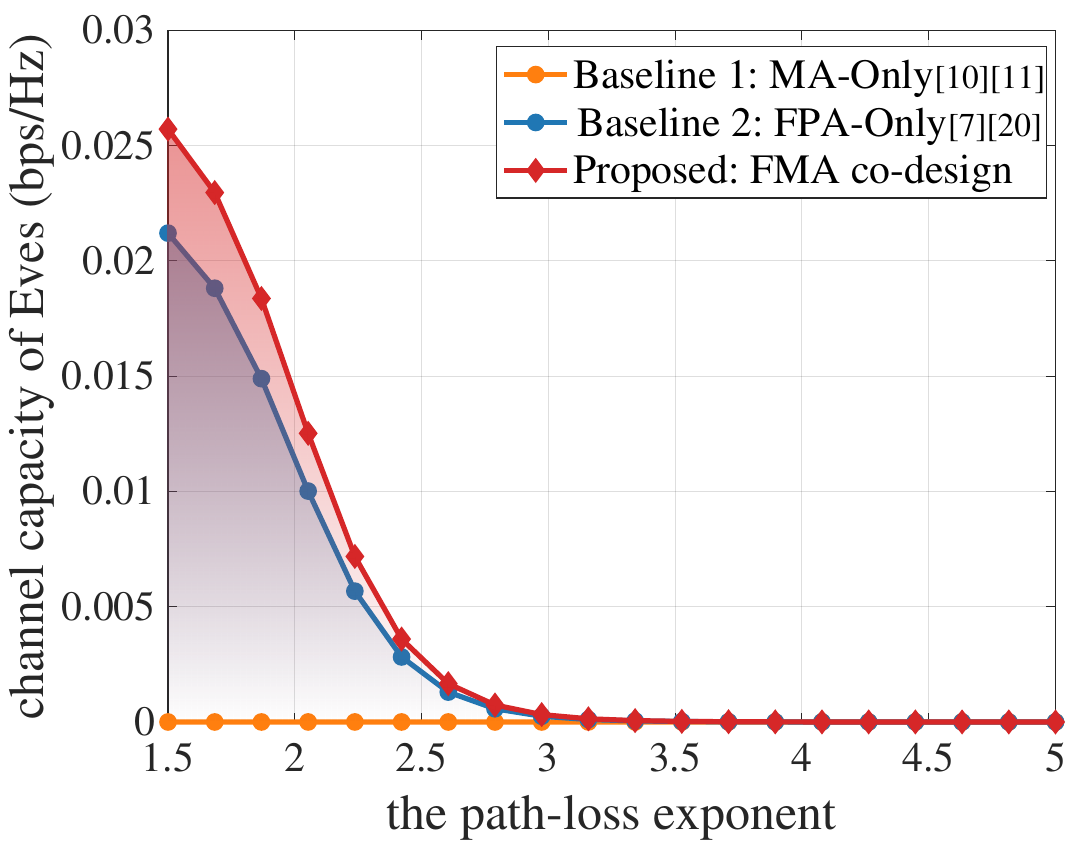}\label{fig:path_loss_b}}
	\subfloat[Secrecy rate vs. path-loss exponent]{\includegraphics[width=2in]{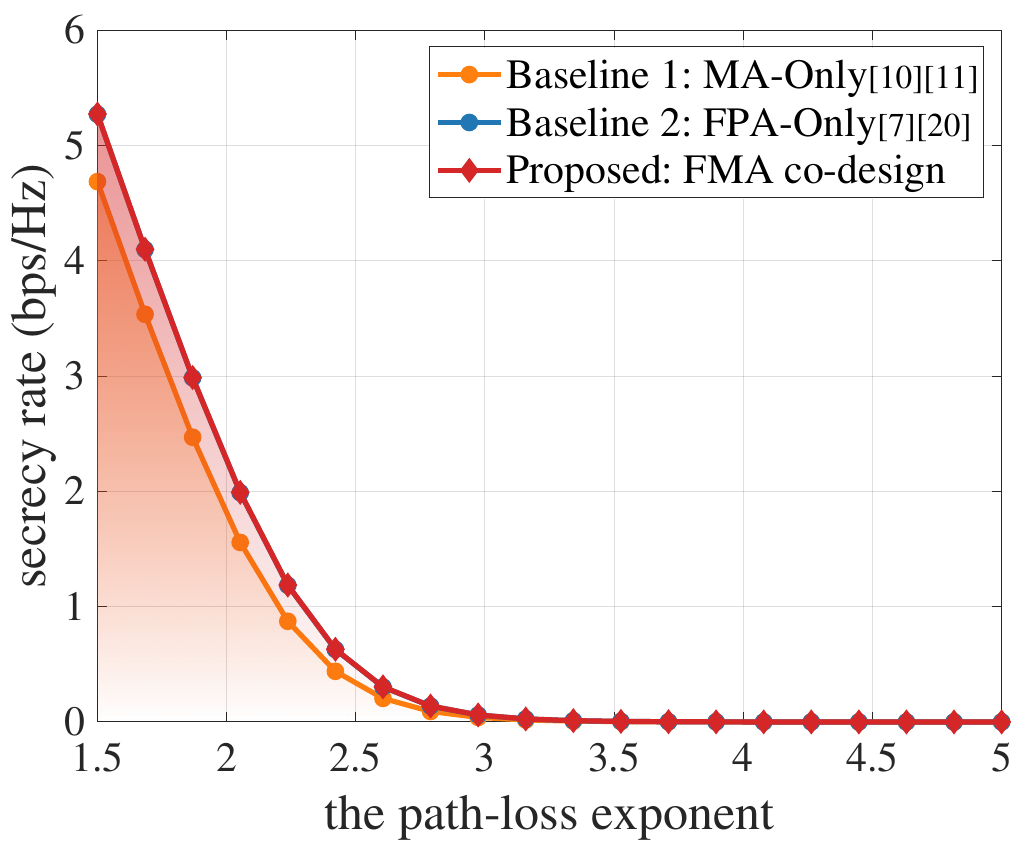}\label{fig:path_loss_c}}
	\caption{\small Channel capacity and secrecy rate for different settings of path-loss exponent.}
	\label{fig:path_loss}
\end{figure*}

Fig. \ref{fig:path_loss} evaluates the joint impact of path-loss exponent and antenna configuration strategies on the Bob's channel capacity, Eve's channel capacity, and secrecy rate. To isolate beamforming and antenna positioning effects, our analysis progresses from idealized to practical channels by incorporating path-loss dynamics, revealing critical performance boundaries under realistic propagation conditions.

As shown in Fig. \ref{fig:path_loss_a}, the \textbf{FMA co-design} framework maintains 0.6bps/Hz and 1.2bps/Hz advantages of Bob's channel capacity over \textbf{MA-Only} baseline for $\alpha=1.5$ and $\alpha=2.8$. This advantage diminishes beyond $\alpha=2.8$ as fundamental propagation limits dominate system behavior.
For Eve's channel capacity shown in Fig. \ref{fig:path_loss_b}, the \textbf{MA-Only} baseline achieves universal eavesdropping suppression, constraining eavesdropping channel capacity below 0.01bps/Hz for $\alpha\geq2$. While the \textbf{FMA co-design} framework exhibits slightly higher initial eavesdropping channel capacity ($0.026$ bps/Hz at $\alpha=1.5$), it reduces this vulnerability by 97.3\% (0.0007bps/Hz) at $\alpha=2.8$. Similarly, the \textbf{MA-Only} baseline exhibits slightly lower initial eavesdropping channel capacity ($0.021$ bps/Hz at $\alpha=1.5$), it reduces this vulnerability by 97.1\% (0.0006bps/Hz) at $\alpha=2.8$.

Finally, as shown in Fig. \ref{fig:path_loss_c}, the \textbf{FMA co-design} framework achieves a peak secrecy rate of 5.3bps/Hz at $\alpha=1.5$, retaining 63.2\% of secrecy rate (3.35bps/Hz) at $\alpha=2.5$-a $1.7\times$ slower decay rate compared to the \textbf{MA-Only} baseline, which varies from 4.7bps/Hz to 0.3bps/Hz over the same $\alpha$ range. All methods collapse to near-zero secrecy rates ($\leq 0.1$bps/Hz) when $\alpha\geq3.0$, marking a critical threshold for the effectiveness of secure communication in severe path-loss environments.

These results quantitatively validate the robustness and effectiveness of our proposed framework under practical channel conditions, while also highlighting the significant effects of path-loss exponent on achievable the system security.

\section{Conclusions and future works} \label{sec:conclusions}
In this paper, through the joint optimization of MA positions, BF matrices for MAs, and BF matrices for fixed antennas in a dynamic vehicular network environment, our proposed scheme effectively balances the improvement of the secrecy rate of legitimate receiving vehicles and controlling power allocation. Using the potential coverage voids that may arise from the movement of MAs, our method achieves the creation of noise nulls at the location of legitimate receiving vehicles. The optimal solution is sought using a momentum-accelerated gradient projection algorithm and an alternating optimization algorithm, further improving the system's efficiency and response speed.
In this paper, we proposed a communication antenna scheme that combines FPA and MA. By jointly optimizing the antenna position and BF vector, high-secrecy rate communication is achieved in the IoV environment. AO and NMPGA algorithms are used to solve the non-convex optimization problem. The simulation results have verified the efficiency and robustness of the scheme. In addition, in future work, we will study MA from two aspects: 1) The current work is based on the assumption of perfect CSI based on known Eves. We will consider the imperfect CSI of Eves and use MA to enable physical layer security. 2) We will explore and build a prototype and physical verification system for MA.

\Acknowledgements{This work is supported by the National Natural Science Foundation of China with Grants 62301076, 62341101, and 62321001, the Macao Young Scholars Program with Grants AM2023015.}

\bibliographystyle{IEEEtran}
\bibliography{Bibliography}
\end{document}